\documentclass[twocolumn]{autart}    
\usepackage{graphics} 
\usepackage{epsfig} 
\usepackage{mathptmx} 
\usepackage{romannum}
\usepackage{algorithm} 
\usepackage{algpseudocode} 
\usepackage{amsfonts}
\usepackage{amsmath}
\usepackage{amssymb}
\DeclareMathOperator*{\argmin}{arg\,min}
\DeclareMathOperator*{\argmax}{arg\,max}
\usepackage{graphicx}      
\usepackage{url}
\usepackage{color}
\usepackage{mathtools}

\begin{document}

\begin{frontmatter}

\title{Robust Action Governor for Uncertain Piecewise Affine Systems \\ with Non-convex Constraints and Safe Reinforcement Learning\thanksref{footnoteinfo}} 

\thanks[footnoteinfo]{This paper was not presented at any IFAC meeting. Corresponding author Yutong Li. The first two authors contributed equally to this work. }

\author[Umich,Ford]{Yutong Li}\ead{yli388@ford.com},    
\author[Umich]{Nan Li}\ead{nanli@umich.edu},               
\author[Ford]{H. Eric Tseng}\ead{htseng@ford.com},  
\author[Umich]{Anouck Girard}\ead{anouck@umich.edu},    
\author[Ford]{Dimitar Filev}\ead{dfilev@ford.com},               
\author[Umich]{Ilya Kolmanovsky}\ead{ilya@umich.edu}  

\address[Umich]{Department of Aerospace Engineering, University of Michigan, Ann Arbor, MI 48105, USA}  
\address[Ford]{Ford Motor Company, Dearborn, MI, 48126 USA}             
 


\begin{abstract}                          
The action governor is an add-on scheme to a nominal control loop that monitors and adjusts the control actions to enforce safety specifications expressed as pointwise-in-time state and control constraints. In this paper, we introduce the Robust Action Governor (RAG) for systems the dynamics of which can be represented using discrete-time Piecewise Affine (PWA) models with both parametric and additive uncertainties and subject to non-convex constraints. We develop the theoretical properties and computational approaches for the RAG. After that, we introduce the use of the RAG for realizing safe Reinforcement Learning (RL), i.e., ensuring all-time constraint satisfaction during online RL exploration-and-exploitation process. This development enables safe real-time evolution of the control policy and adaptation to changes in the operating environment and system parameters (due to aging, damage, etc.). We illustrate the effectiveness of the RAG in constraint enforcement and safe RL using the RAG by considering their applications to a soft-landing problem of a mass-spring-damper system.
\end{abstract}

\end{frontmatter}

\section{Introduction}
Cyber-physical systems (CPS) are becoming increasingly adopted in a broad range of applications. Examples of CPS include self-driving cars \cite{litman2020autonomous}, manufacturing systems \cite{hashemi2020operations}, and medical devices \cite{lee2010medical}. Piecewise-affine (PWA) models are capable of representing the interaction of continuous and discrete dynamics, which frequently arise in CPS, and, thereby, have gained increasing research interest \cite{sontag1981,heemels2001equivalence,borrelli2017predictive}.

Safety is a crucial consideration for many CPS. Many safety requirements can be expressed as state and control constraints. One route to handle safety constraints is through incorporating them into the controller design. Model predictive control (MPC) is a popular approach due to its ability to explicitly enforce constraints \cite{camacho2010model,borrelli2017predictive}. MPC for PWA models has been studied in, e.g., \cite{lazar2006stabilizing} and \cite{borrelli2005dynamic}. 
Another route to handle constraints is through add-on, supervisory schemes that complement the existing/legacy controller. The reference governor (RG) is one of such add-on schemes, which enforces constraints by monitoring and manipulating the reference inputs to the closed-loop system \cite{garone2017reference}. RG design strategies based on PWA models are presented in \cite{borrelli2009reference}. The control barrier function (CBF) provides another way to handle constraints via a supervisory scheme \cite{wieland2007constructive}: A CBF and a Control Lyapunov Function (CLF) are integrated into a quadratic program (QP) to promote stability and enforce constraints \cite{ames2016control}.

Enforcing state and control constraints is more challenging when there exist discrepancies between the real-world system and its control-oriented model. Such discrepancies may arise from unknown model parameters, unmodeled dynamics, external disturbances, etc. These are collectively referred to as {\it uncertainties}. 
\begin{figure}[thpb]
      \centering
      \includegraphics[scale=0.4]{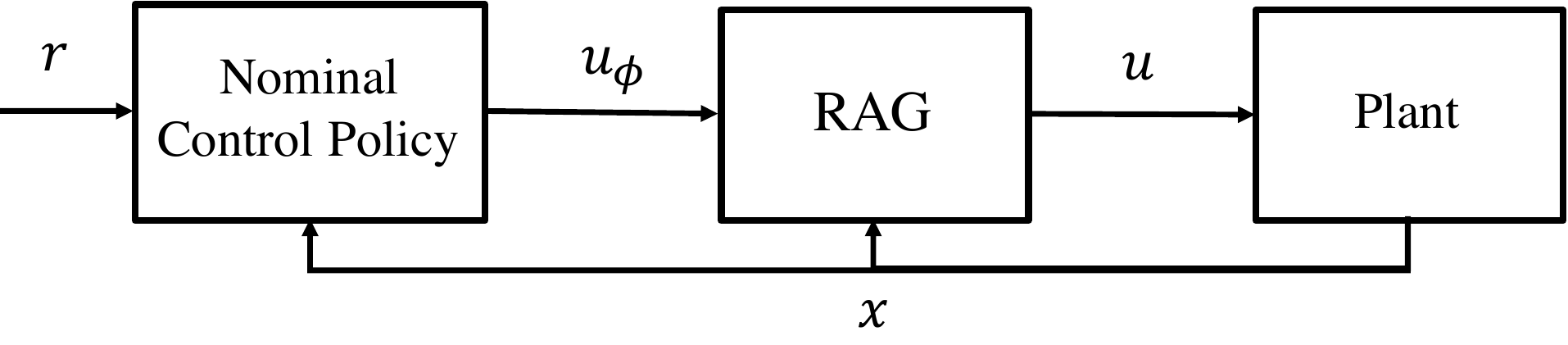}
      \caption{Schematic diagram of the RAG.}
      \label{fig:RAG_concept}
\end{figure}

Robust model predictive control (RMPC) is a variant of MPC for addressing uncertainties \cite{bemporad1999robust}. A popular RMPC approach is the tube MPC, where uncertain trajectories are bounded by a tube and constraints are tightened so that enforcing the tightened constraints for a nominal trajectory implies constraint satisfaction by all trajectories admissible under the bounded uncertainty. Another approach is the min-max MPC, where control performance under the worst-case uncertainty is optimized using a min-max approach. Typical approaches based on RMPC require the (re-)design of the controller. Robust versions of RG and CBF to enforce constraints in the presence of uncertainties have also been proposed, for linear systems in \cite{kolmanovsky1998theory} and for nonlinear systems with known input-to-state stability (ISS) Lyapunov functions in \cite{gilbert2002nonlinear,jankovic_robust_2018,AmesRobust2020}.

In this paper, we focus on another add-on, supervisory scheme, called the Robust Action Governor (RAG), for handling safety constraints of CPS. The RAG is a scheme which monitors and modifies the output of the nominal control policy, as illustrated in Fig.\ref{fig:RAG_concept}. The RAG monitors the control action generated by the nominal policy and, when necessary, minimally modifies it to ensure that the plant response to the modified control action satisfies prescribed constraints. The action governor (AG) scheme was first introduced for linear systems in \cite{li2020action} and then extended to PWA models subject to only additive disturbances in \cite{li2021robust}. In this paper, we further develop the AG theory and computational approaches to enable AG design based on PWA models with both parametric and additive uncertainties. 

More specifically, this paper differentiates itself from our previous work in \cite{li2021robust} and the work of \cite{schurmann2020set} by considering both parametric and additive uncertainties in the PWA model, while \cite{li2021robust} and \cite{schurmann2020set} only consider additive disturbances. On the one hand, incorporating both parametric and additive uncertainties significantly enlarges the class of systems to which the RAG can be applied. This is because many uncertainties, such as uncertainties in system parameters, can be straightforwardly represented as parametric uncertainties in the model, but can be difficult, if not impossible, to be non-conservatively account for with additive disturbances \cite{liu2019improving,bhattacharyya2017robust}. On the other hand, in order to address parametric uncertainties, new methods for computation and analysis of the RAG are needed, which necessitates the developments in this paper, including the new proofs for the RAG properties.

We note that the RAG addressed in this paper focuses on enforcing non-convex constraints, i.e., the region within which the system state needs to be maintained to ensure safety is generally non-convex. Such as in the obstacle avoidance problems frequently encountered in autonomous vehicle/mobile robot applications \cite{funke2016collision,verginis2021adaptive}. Many previous approaches to PWA systems, including the RMPC approach in \cite{lazar2004stabilization}, the approaches based on RG in \cite{borrelli2009reference} and based on CBF in \cite{taylor2020adaptive,prajna2004safety}, have focused on convex constraints. Although non-convex constraints are addressed in \cite{rakovic2007robust,nicotra2019spacecraft,romagnoli2019feedback,notomista2021safety}, by either RG or CBF, these references do not deal with PWA systems, which is the focus of this work. Furthermore, to the best of our knowledge, no methods using CBFs to treat PWA models with both parametric and additive uncertainties currently exist, partly because CBF for a PWA system with parametric uncertainties is not easy to compute \cite{biswas2005survey}. Some other differences of the AG approach as compared to the RG and the CBF approaches for general systems are highlighted in \cite{li2020action,li2021robust}.

In this paper, after introducing the RAG and describing its properties, we illustrate its use for safe RL. Specifically, we propose a safe RL framework where the RAG is used as a safety supervisor to monitor the actions selected by a nominal RL agent and correct the unsafe actions to safe ones. This way, one can augment an arbitrary RL algorithm with the RAG after which the combined algorithm ensures all-time safety during the entire learning process. Along these lines, in this paper we demonstrate the integration of Q-learning with the RAG to realize safe Q-learning. In addition, we demonstrate that the development of an explicit safe RL policy to reduce the online computational cost of the safe RL is feasible using the imitation learning. Although similar approaches have been proposed in the model-based RL literature  \cite{aswani2013,fisac2018,larsen2017,sloth2012,li2018safe}, none of these references deals with PWA systems that have both parametric and additive uncertainties.

The main contributions of this paper are as follows: 
\begin{itemize}
  \item We extend the RAG scheme developed in \cite{li2021RLRAG,li2021robust} to PWA systems subject to both additive and parametric uncertainties. This is the first time that the parametric uncertainties are accommodated in the RAG design.  
  \item We analyze theoretical properties of the proposed RAG scheme, including robust constraint satisfaction and recursive feasibility. New safe set computational approaches that handle parametric uncertainties are developed. These theoretical properties and computational approaches are based on the PWA system model subject to non-convex constraints and both additive and parametric uncertainties, which distinguish our RAG scheme versus existing robust control techniques. For instance, while robust variants of the CBF have been proposed, finding CBFs and CLFs for PWA systems is nontrivial \cite{biswas2005survey}, especially in the presence of both additive and parametric uncertainties.  
  \item We extend the safe RL framework in \cite{li2021RLRAG} based on the proposed RAG scheme. Both additive and parametric uncertainties are taken into account within the safe exploration region design for the RL agent. Moreover, in order to reduce the online computational cost, an explicit safe RL policy is obtained by using imitation learning.
  \item We demonstrate the effectiveness of the proposed RAG scheme and safe RL framework using an example of soft-landing control for a mass-spring-damper system. Such an example is relevant to applications in the ones of aircraft flight control \cite{rajaram2017laplace}, robotic manipulation \cite{kraus2020mechatronic,nguyen2020active} and cooperative navigation \cite{urcola2008cooperative}. 
\end{itemize}

The remainder of the paper is organized as follows: In Section \ref{sec:2} the main assumptions and problem setup are introduced. In Section \ref{sec:3} the main
theoretical results and computational approaches of the RAG are presented. In Section \ref{sec:4} the safe RL framework is introduced. Numerical examples and results are reported in Section \ref{sec:5}.  Section \ref{sec:6} contains concluding remarks and directions
for future work.

\textit{Definitions and Notation:} Given two sets $A$ and $B$, then $A\oplus B=\{a+b:a\in A,b\in B\}$ (Minkowski sum), and $A\sim B=\{a:a\oplus B\subseteq A\}$ (Pontryagin difference). $\text{Proj}_x(A)$ represents the projection of the set $A$ onto $x$ space. A polyhedron (polyhedral set) is the intersection of a finite number of closed and/or open halfspaces, a polytope is a closed and bounded (equivalently, compact) polyhedron and a polygon is the union of a finite number of polyhedra (and is thus not necessarily convex). 
\section{Problem formulation}\label{sec:2}

In this paper, we focus on systems that can be described by the following PWA model,
\begin{align}\label{equ:PWASysDynParaDis_1}
\begin{split}
   x(t+1) &= f_{q(t)}(x(t),u(t),w(t)) \\ 
   &= A_{q(t)}(w^\text{p}(t))x(t)+B_{q(t)}(w^\text{p}(t))u(t) \\
   &\quad + f_{q(t)}(w^\text{p}(t)) + E_{q(t)}(w^\text{p}(t)) w^\text{a}(t), 
\end{split}
\end{align}
where $x(t)\in \mathbb{R}^n$ represents the system state at the discrete time instant $t\in \mathbb{Z}_{\geq 0}$, $u(t)$ represents the control input, $w(t) = (w^\text{p}(t), w^\text{a}(t))$ represents unmeasured uncertainties, including both parametric uncertainty $w^\text{p}(t)$ and additive uncertainty $w^\text{a}(t)$, and $q(t) \in Q = \{1,2,...,n_q\}$ represents the current mode of the system. We assume the mode $q(t)$ is state-dependent, in particular,
\begin{equation}\label{equ:PWASysDynParaDis_10}
q(t) = q \quad \text{if} \quad x(t) \in P_q,
\end{equation}
where $P_q \subset \mathbb{R}^n$ is a polyhedral set and the collection $\{P_q\}_{q = 1}^{n_q}$ forms a partition of $\mathbb{R}^n$. This means the dynamics \eqref{equ:PWASysDynParaDis_1} hold for all $x \in P_{q(t)}$. Given a mode $q \in Q$, we assume the uncertain matrices $(A_q(w^\text{p}),B_q(w^\text{p}),f_q(w^\text{p}),E_q(w^\text{p}))$ can be parameterized by $w^\text{p}$ as
\begin{align}\label{equ:PWASysDynParaDis_2}
\begin{split}
   & A_q(w^\text{p}) = \sum_{j=1}^{n^\text{p}_{q}} w^\text{p}_{j} A_{q,j},\,\, B_q(w^\text{p}) = \sum_{j=1}^{n^\text{p}_{q}} w^\text{p}_{j} B_{q,j}, \\
   & f_q(w^\text{p}) = \sum_{j=1}^{n^\text{p}_{q}} w^\text{p}_{j} f_{q,j},\,\, E_q(w^\text{p}) = \sum_{j=1}^{n^\text{p}_{q}} w^\text{p}_{j} E_{q,j},
\end{split}
\end{align}
where $(A_{q,j},B_{q,j},f_{q,j},E_{q,j})$, for $j = 1,...,n^\text{p}_{q}$, are known constant matrices. Given the current mode $q(t) \in Q$, we assume the uncertainties $w(t)$ take values according to
\begin{equation}
    w(t) = (w^\text{p}(t), w^\text{a}(t)) \in W^\text{p}_{q(t)} \times  W^\text{a}_{q(t)} = W_{q(t)}, 
\end{equation}
where $W^\text{a}_{q(t)}$ is a mode-dependent polytope, and $W^\text{p}_{q(t)}$ is a mode-dependent unit simplex, written as
\begin{equation}\label{equ:simplex}
    W^\text{p}_{q(t)} = \{w^\text{p} \in \mathbb{R}^{n^\text{p}_{q(t)}}: \sum_{j=1}^{n^\text{p}_{q(t)}} w^\text{p}_{j} = 1,\  w^\text{p}_{j} \geq 0\}.
\end{equation}
We note that the above equations, \eqref{equ:PWASysDynParaDis_1}-\eqref{equ:simplex}, define a state‐dependent switching system with both parametric and additive uncertainties. 

We now consider the following safety requirements,
\begin{equation}\label{equ:Safety_1}
   x(t) \in X, \quad u(t) \in U_{q(t)}, \quad \forall t \in \mathbb{Z}_{\geq 0},
\end{equation}
where $X \subset \mathbb{R}^n$ is a mode-independent safe region for the state, and $U_{q(t)}$ is a mode-dependent safe region for the control input and is assumed to be a polytope. In particular, we consider nonconvex safety requirements in the form of
\begin{equation}\label{equ:ExcluZone}
   x(t) \in X = \bigcup_{i = 1}^{r_0} X^i,
\end{equation}
where $X^i$ is a polyhedron for each $i=1,...,r_0$, and thus $X$ is a polygon (not necessarily convex). 

Oftentimes, it is easier to design a nominal control policy that achieves liveness properties such as stabilization and reference tracking than to develop one that ensures both liveness and the strict satisfaction of safety requirements. For instance, for linear systems, a variety of tools exist for stabilization and reference tracking, such as pole placement, linear-quadratic regulator, etc., but these conventional tools cannot strictly handle constraints. Along these lines, suppose a nominal control policy, $\phi$, has been defined for the system \eqref{equ:PWASysDynParaDis_1}, 
\begin{equation}\label{equ:nominalCtr}
   u_{\phi}(t) = \phi(x(t),r(t),t),
\end{equation}
where $r(t) \in \mathbb{R}^r$ denotes a reference command signal corresponding to the nominal control objective (such as tracking). We note that there is no further assumption on the nominal control policy $\phi$ and it can be nonlinear and/or time-varying. For instance, $\phi$ may represent a neural network controller that evolves over time via online learning, which will be further discussed in Sections~\ref{sec:4} and \ref{sec:5}. Note also that for the PWA system \eqref{equ:PWASysDynParaDis_1}, such a control policy is typically mode-dependent. However, since the mode $q(t)$ is determined by the state $x(t)$ according to \eqref{equ:PWASysDynParaDis_10}, the dependence on $q(t)$ eventually becomes the dependence on $x(t)$. So we have chosen not to include $q(t)$ as an input to the control policy $\phi$ in \eqref{equ:nominalCtr}.

In this paper, we consider a scenario where the safety requirement \eqref{equ:ExcluZone} has not been incorporated or cannot be guaranteed by the nominal policy \eqref{equ:nominalCtr}. The objective of this paper is to develop an add-on control scheme for the system \eqref{equ:PWASysDynParaDis_1} to strictly handle the safety requirements \eqref{equ:Safety_1}-\eqref{equ:ExcluZone} in the presence of parametric and additive uncertainties.


\section{Robust action governor}\label{sec:3}
The solution we propose is an add-on, supervisory scheme, called the Robust Action Governor (RAG), illustrated in Fig.~\ref{fig:RAG_concept}. The RAG monitors the nominal control input $u_{\phi}$ and, if necessary, minimally modifies $u_{\phi}$ to ensure that the system state $x$ stays inside $X$ at all time instants even in the presence of model uncertainties/disturbances.

In particular, the RAG operates by solving the following constrained optimization problem at each time instant $t$ under the assumption that the current state $x(t)$ and mode $q(t)$ are both known, 
\begin{align}\label{equ:RAG}
\begin{split}
    & u(t)  \in\argmin_{u \in U_{q(t)}}\ {\|u-u_{\phi}(t)\|}_S^2 \\
    &\text{ s.t. }\, A_{q(t)}(w^\text{p}_{q(t)}(t))x(t)+B_{q(t)}(w^\text{p}_{q(t)}(t))u + f_{q(t)}(w^\text{p}_{q(t)}(t))\\
    &\quad\quad+E_{q(t)}(w^\text{p}_{q(t)}(t))w^\text{a}_{q(t)}(t)\in \chi_{\text{safe}},\ \forall w^\text{p}_{q(t)}(t) \in W^\text{p}_{q(t)},\\&\quad\quad \forall w^\text{a}_{q(t)}(t) \in W^\text{a}_{q(t)},
\end{split}
\end{align}
where $\chi_{\text{safe}}\subset X$ is a ``safe set'' which will be introduced in the next section. The function $\| \cdot \|_S = \sqrt{(\cdot)^T S (\cdot)}$, where $S$ is a positive-definite penalty matrix, is used to promote minimization of the difference between the nominal control $u_{\phi}$ and the modified control $u$. 

\subsection{Safe set}\label{sec:31}
To ensure robust satisfaction of the non-convex safety requirement \eqref{equ:ExcluZone} at all future time instants $t$, the safe set $\chi_{\text{safe}}$ should have the following property: Suppose $x(0) \in \chi_{\text{safe}}$, then there exists a state-feedback control law, $u(x)$, such that 
\begin{itemize}
  \item the control inputs $u(t) = u(x(t))$ take values necessarily in $U_{q(t)}$ for all $t = 0,1,...$
  \item under $u(x)$, the states $x(t)$ are necessarily inside $X$ for all $t = 1,2,...$
\end{itemize}
under any uncertainty realizations $\{w(0),w(1),...\}$ that satisfy $w(t) \in  W_{q(t)}$ for all $t = 0,1,...$

We first introduce the following lemma to facilitate the definition of $\chi_\text{safe}$.

\textbf{Lemma 1}
\textit{Let $J$ be an affine function of $w \in W$, $W$ be a polytope with vertices $\{\bar{w}_i\}_{i=1}^{n_W}$, and $Z$ be a polyhedron. Then,
\begin{equation} 
    J(w) z \in Z, \quad \forall w\in W, 
\end{equation}
if and only if 
\begin{equation} 
    J(\bar{w}_i)z \in Z, \quad i=1,...,n_W.
\end{equation}}

\textit{Proof:} The proof follows from Lemma 11.1 of \cite{borrelli2017predictive}.
\hfill$\blacksquare$


To define $\chi_{\text{safe}}$, we first consider a sequence of sets, $\chi_{\text{safe},k}$, defined recursively as follows,
\begin{align} \label{equ:SafeSet0}
\begin{split}
\chi_{\text{safe},0}=X,
\end{split}
\end{align}
and for $k\in \mathbb{Z}_{\geq 1}$ 
\begin{subequations}\label{equ:SafeSet}
\begin{align} \label{equ:SafeSet1}
    &\chi_{\text{safe},k} =\text{Proj}_x({\Lambda_{\text{safe},k-1}})
\end{align}
where
\begin{align}\label{equ:SafeSet2}
    &\Lambda_{\text{safe},k}\nonumber\\
     &=\bigcup_{q=1}^{n_q}\bigcup_{i=1}^{r_{k}}\{(x,u):x\in \chi_{\text{safe},k}\cap{P}_q, u\in U_q,\text{s.t.}\, \forall w\in W_q,\nonumber\\&
     A_q(w^\text{p})x+B_q(w^\text{p})u+f_q(w^\text{p})+E_q(w^\text{p})w^\text{a}\in \chi_{\text{safe},k}^i\}\nonumber\\
     &=\bigcup_{q=1}^{n_q}\bigcup_{i=1}^{r_{k}}\{(x,u): x\in \chi_{\text{safe},k}\cap{P}_q,u\in U_q,\text{s.t.}\, \forall w^\text{a}\in W_q^\text{a},\nonumber\\&
     A_q(w^\text{p})x+B_q(w^\text{p})u+f_q(w^\text{p})+E_q(w^\text{p})w^\text{a}\in \chi^i_{\text{safe},k},\forall w^\text{p} \in W^\text{p}_{q}\}\nonumber\\
     &=\bigcup_{q=1}^{n_q}\bigcup_{i=1}^{r_{k}}\{(x,u): x\in \chi_{\text{safe},k}\cap{P}_q,u\in U_q,\text{s.t.}\, \forall w^\text{a}\in W_q^\text{a},\nonumber\\&
     A_{q,j}x+B_{q,j}u+f_{q,j}+E_{q,j}w^\text{a}\in \chi^i_{\text{safe},k},j=1,2,...,n^\text{p}_q\}\nonumber\\
    &=\bigcup_{q=1}^{n_q}\bigcup_{i=1}^{r_{k}}\{(x,u): x\in \chi_{\text{safe},k}\cap{P}_q,u\in U_q,\text{s.t.}\, \nonumber\\&j=1,2,...,n^\text{p}_q,
     A_{q,j}x+B_{q,j}u\in \chi^i_{\text{safe},k}\sim E_{q,j}W_q^\text{a}-f_{q,j}\}\nonumber\\
    &=\bigcup_{q=1}^{n_q}\bigcup_{i=1}^{r_{k}}\{(x,u): x\in \chi_{\text{safe},k}\cap{P}_q,u\in U_q,\text{s.t.} \nonumber\\&
      \begin{bmatrix}
      \bar{A}_q & \bar{B}_q
     \end{bmatrix}
     \begin{bmatrix}
      x \\ u
     \end{bmatrix}\in
     \prod_{j=1}^{n^\text{p}_q} (\chi^i_{\text{safe},k}\sim E_{q,j}W_q^\text{a}-f_{q,j}) \}.
\end{align}
\end{subequations}
where $\chi_{\text{safe},k}$ is assumed to be a polygon, written as $\chi_{\text{safe},k} =\bigcup_{i=1}^{r_{k}}\chi^i_{\text{safe},k}$ with polyhedra $\chi^i_{\text{safe},k}$, $i = 1,...,r_{k}$. $\prod$ represents the Cartesian product where $\prod_{j=1}^s S_j=\{(x_1,...,x_s):x_j\in S_j,\forall j\in\{1,...,s\}\}$, and $\begin{bmatrix}
 \bar{A}_q & \bar{B}_q
\end{bmatrix}$ are defined as
\begin{align}
\begin{split}
   & \begin{bmatrix}
    \bar{A}_q & \bar{B}_q
   \end{bmatrix} = \begin{bmatrix}
    {A}_{q,1} & {B}_{q,1}\\
    \vdots & \vdots\\
    {A}_{q,n_q^{\text{p}}} & {B}_{q,n_q^{\text{p}}}
   \end{bmatrix}.
\end{split}
\end{align}
Note that in \eqref{equ:SafeSet2}, between the third line and the forth line, we use the result from Lemma~1. In what follows we first show that if $\chi_{\text{safe},k-1}$ is indeed a polygon, then $\chi_{\text{safe},k}$ defined according to \eqref{equ:SafeSet} is also a polygon. A direct consequence of this fact is that if $\chi_{\text{safe},0} = X$ is a polygon, then the sequence of sets, $\{\chi_{\text{safe},k}\}_{k = 0}^{\infty}$, defined recursively according to \eqref{equ:SafeSet} are all polygons. Then, it can be shown that if $x(0)\in \chi_{\text{safe},k}$, $k\in \mathbb{Z}_{\geq 0}$, then it is possible to keep $x(t)$ within $X$ for at least $k$ steps. These properties are elaborated in the following propositions.

\textbf{Proposition 2}
\textit{If }$\chi_{\text{safe},k-1} =\bigcup_{i=1}^{r_{k-1}}\chi^i_{\text{safe},k-1}$, \textit{for some }$k\in\mathbb{Z}_{\geq 1}$, \textit{is indeed a polygon with polyhedra }$\chi^i_{\text{safe},k-1}$, $j = 1,...,r_{k-1}$,\textit{ then }$\chi_{\text{safe},k}$ \textit{defined according to \eqref{equ:SafeSet} is also a polygon.}

\textit{Proof:} 
Let $\chi_{\text{safe},k-1}=\bigcup_{i=1}^{r_{k-1}}\chi^i_{\text{safe},k-1}$ be a polygon for some $k\in\mathbb{Z}_{\geq 1}$, recall that the Pontryagin difference of a polyhedron and a polytope is a polyhedron, and the Cartesian product of polyhedron is a (higher-dimensional) polyhedron, thus $\prod_{j=1}^{n^\text{p}_q} (\chi^i_{\text{safe},k-1}\sim E_{q,j}W_q^\text{a}-f_{q,j})$ is a polyhedron. Furthermore, because the preimage of a polyhedron under a linear transformation is a polyhedron, $\chi_{\text{safe},k-1}\cap P_q$ is a polygon, $U_q$ is a polytope, and the intersection of a polygon and a polytope is a polygon, hence $\{(x,u): x\in \chi_{\text{safe},k-1}\cap{P}_q,u\in U_q,\text{s.t.}
      \begin{bmatrix}
      \bar{A}_q & \bar{B}_q
     \end{bmatrix}
     \begin{bmatrix}
      x \\ u
     \end{bmatrix}\in
     \prod_{j=1}^{n^\text{p}_q} (\chi^i_{\text{safe},k-1}\sim E_{q,j}W_q^\text{a}-f_{q,j})\}$ is a polygon, so as $\Lambda_{\text{safe},k-1}$. Finally, as the projection of a polygon onto a subspace is a (lower-dimensional) polygon \cite{rakovic2006reachability}, and according to \eqref{equ:SafeSet1}, we have $\chi_{\text{safe},k}=\text{Proj}_x(\Lambda_{\text{safe},k-1})$ is a polygon.  
\hfill$\blacksquare$ 

\textbf{Proposition 3}
\textit{If} $x(0) \in \chi_{\text{safe},k}$\textit{, then there exists a state-feedback control sequence $\{u_0(x(0)),...,u_{k-1}(x(k-1))\} \in U_{q(0)} \times \cdots \times U_{q(k-1)}$, such that for any disturbance sequence $\{w(0),...,w(k-1)\} \in W_{q(0)} \times \cdots \times W_{q(k-1)}$, $x(j) \in X$ for all $0 \leq j \leq k$.}

\textit{Proof:} The proof is by induction. 
For $k = 1$, $x(0) \in \chi_{\text{safe},1}$ implies that there exists a control $u(x(0))$, such that $(x(0),u(x(0))) \in \Lambda_{\text{safe},0}$ from \eqref{equ:SafeSet1}. Thus, by the definition of $\Lambda_{\text{safe},0}$, we have that $x(0)\in \chi_{\text{safe},0}=X$, and there exist $q_0\in \{1,...,n_q\}$ and $i_0\in\{1,...,r_0\}$, such that there exists $u_0(x(0)) \in U_{q_0}$, for all $w_0 \in W_{q_0}$, $x(1)=f_{q_0}(x(0),u_0,w_0) \in \chi^{i_0}_{\text{safe},0} \subset \bigcup_{i=1}^{r_0}\chi^i_{\text{safe},0} = \chi_{\text{safe},0}=X$. Suppose the statement has been proven for all $\chi_{\text{safe},j}$, $1 \leq j \leq k$. If $x(0) \in \chi_{\text{safe},k+1}$, then from \eqref{equ:SafeSet1} we have that there exists $u(x(0))$, such that $(x(0),u(x(0)))\in \Lambda_{\text{safe},k}$. Again, from the definition of $\Lambda_{\text{safe},k+1}$ in \eqref{equ:SafeSet2}, there exist $q_0\in \{1,...,n_q\}$ and $i_0\in\{1,...,r_k\}$ such that there exists $u_0(x(0)) \in U_{q_0}$, for all $w_0 \in W_{q_0}$, $x(1)=f_{q_0}(x(0),u_0,w_0) \in \chi^{i_0}_{\text{safe},k}\subset \bigcup_{i=1}^{r_k}\chi^{i}_{\text{safe},k} = \chi_{\text{safe},k}$. Since $x(1) \in \chi_{\text{safe},k}$, by our induction hypothesis above, there exists $\{u_1(x(1)),...,u_k(x(k))\} \in U_{q(1)} \times \cdots \times U_{q(k)}$, such that for all $\{w_1,...,w_k\} \in W_{q(1)} \times \cdots \times W_{q(k)}$, $x(j+1) \in X$ for all $0 \leq j \leq k$. That is, there exists $\{u_0(x(0)),u_1(x(1)),...,u_k(x(k))\} \in U_{q(0)} \times U_{q(1)} \times \cdots \times U_{q(k)}$, such that for all $\{w_0,w_1,...,w_k\} \in W_{q(0)} \times W_{q(1)}\times \cdots \times W_{q(k)}$, $x(j+1) \in X$, for $0 \leq j+1 \leq k+1$. This proves the statement for $k+1$ and hence completes the induction step.  
\hfill$\blacksquare$

\textbf{Proposition~4} \textit{For each $k = 0,1,...$, it holds that} $\chi_{\text{safe},k+1} \subset \chi_{\text{safe},k}$, \textit{i.e.,} $\chi_{\text{safe},k}$ \textit{is a decreasing sequence of sets. Thereby,} $\chi_{\text{safe},\infty} = \lim_{k \to \infty} \chi_{\text{safe},k}$ \textit{exists\footnote{In the set-theoretic sense \cite{rudin1991functional}.} and satisfies }$\chi_{\text{safe},\infty} \subset \chi_{\text{safe},k}$ \textit{for all $k$.}

\textit{Proof:} Suppose $x\in \chi_{\text{safe},k+1}$ for some $k\in \mathbb{Z}_{\geq 0}$. According to \eqref{equ:SafeSet1}, there exists $u(x)$ such that $(x,u(x))\in \Lambda_{\text{safe},k}$. Then, by the definition \eqref{equ:SafeSet2}, there exist $q_k\in \{1,...,n_q\}$ and $i_k\in\{1,...,r_k\}$ such that $x\in \chi_{\text{safe},k}$. This implies $\chi_{\text{safe},k+1}\subset \chi_{\text{safe},k}$, and thus $\chi_{\text{safe},k}$ is decreasing. Thus, the sequence of sets $\chi_{\text{safe},k}$ converges and the limit exists \cite{rudin1991functional}.\hfill $\blacksquare$

On the basis of Proposition 4, we now define $\chi_{\text{safe}}$ as $\chi_{\text{safe}} = \lim_{k \to \infty} \chi_{\text{safe},k}=\bigcap_{k=0}^ \infty \chi_{\text{safe},k}$. 
With this definition, we have the following propositions, which verify that the desired property stated at the beginning of Section~\ref{sec:31} is fulfilled by $\chi_{\text{safe}}$.



\textbf{Proposition 5} \textit{Suppose at a given time instant $t\in\mathbb{Z}_{\geq 0}$, \eqref{equ:RAG} is feasible and $u(t)$ is determined as the solution of \eqref{equ:RAG}, then at $t+1$, $x(t+1) \in X$.}

\textit{Proof:} As \eqref{equ:RAG} is feasible and $u(t)$ is determined as the solution of \eqref{equ:RAG} at $t$, we have $x(t+1)\in \chi_{\text{safe}}$ as the constraint of \eqref{equ:RAG} is satisfied. Additionally, $\chi_{\text{safe}}=\bigcap_{k = 0}^{\infty} \chi_{\text{safe},k}\subset \chi_{\text{safe},0}= X$, thus $x(t+1)\in X$.\hfill $\blacksquare$

\textbf{Proposition 6} \textit{Suppose} $x\in \chi_{\text{safe}}$ \textit{and the following condition holds:} 
\begin{align} \label{equ:P5_1}
\begin{split}
     \text{Proj}_x\Bigg[\bigcap_{k=0}^\infty \Lambda_{\text{safe},k}\Bigg]=\bigcap_{k=0}^\infty \chi_{\text{safe},k}.
\end{split}
\end{align}
\textit{Then,
 there exists $u \in U_{q(x)}$ such that}  $f_{q(x)}(x,u,w) \in \chi_{\text{safe}}$ \textit{for any $w\in W_{q(x)}$.}

\textit{Proof:} Given $x\in \chi_{\text{safe}}=\bigcap_{k=0}^\infty \chi_{\text{safe},k}$, suppose \eqref{equ:P5_1} holds, then there exists a control $u(x)$, such that $(x,u(x))\in \bigcap_{k=0}^\infty \Lambda_{\text{safe},k}$, i.e., $(x,u(x))\in \Lambda_{\text{safe},k}$ for all $k\in\mathbb{Z}_{\geq 0}$. For each $k$, by the definition \eqref{equ:SafeSet2}, there exists $i_k\in\{1,...,r_k\}$ such that $x\in P_{q(x)}$, $u(x)\in U_{q(x)}$, and $f_{q(x)}(x,u(x),W_{q(x)})\subset \chi^{i_k}_{\text{safe},k}\subset \bigcup_{i=1}^{r_{k}}\chi^i_{\text{safe},k}=\chi_{\text{safe},k}$. Since this holds for all $k\in\mathbb{Z}_{\geq 0}$, this implies $f_{q(x)}(x,u(x),W_{q(x)})\subset \bigcap_{k=0}^\infty \chi_{\text{safe},k}=\chi_{\text{safe}}$.
\hfill$\blacksquare$

\textbf{Corollary 7} \textit{Suppose \eqref{equ:P5_1} holds,} $x(0) \in \chi_{\text{safe}}$ \textit{and the RAG operates based on \eqref{equ:RAG}, then \eqref{equ:RAG} is recursively feasible at all $t \in \mathbb{Z}_{\geq 0}$ and the safety requirements in \eqref{equ:Safety_1} are satisfied for all $t \in \mathbb{Z}_{\geq 0}$.}

\textit{Proof:} This result follows directly from Propositions~4 and~5.\hfill$\blacksquare$

Note that \eqref{equ:P5_1} does not always hold (in general, $\text{Proj}_x[\bigcap_{k=0}^\infty \Lambda_{\text{safe},k}]\\\subset\bigcap_{k=0}^\infty \chi_{\text{safe},k}$), it does hold in many cases. The following proposition classifies the circumstances where \eqref{equ:P5_1} is guaranteed to hold.

\textbf{Proposition 8}
\textit{ Suppose there exists $k_0$ such that the sets} $\Lambda_{\text{safe},k}$ \textit{are nonempty and compact for all $k\geq k_0$, then} $\text{Proj}_x\Bigg[\bigcap_{k=0}^\infty \Lambda_{\text{safe},k}\Bigg]=\bigcap_{k=0}^\infty \chi_{\text{safe},k}$.

\textit{Proof:} The proof follows from Proposition~4 in \cite{bertsekas1972infinite}.
\hfill$\blacksquare$

Note that the assumption in Proposition 8 holds in many examples, e.g., continuous systems with additive disturbances, where $X$, $U$ and $W$ are compact sets. See \cite{bertsekas1972infinite} for more examples.

\subsection{Offline and online computations}\label{sec:32}
According to the definition of $\chi_{\text{safe}} = \bigcap_{k=0}^ \infty \chi_{\text{safe},k}$, the exact determination of $\chi_\text{safe}$ relies on the iterative computation of set $\chi_{\text{safe},k}$ according to \eqref{equ:SafeSet} with $k \to \infty$. In practice, as $\chi_{\text{safe},k}$ is a decreasing sequence of sets, we will use a compact approximation of $\chi_{\text{safe}}$ in the RAG, which is denoted as $\tilde{\chi}_{\text{safe},k'}$, where $k'$ being sufficiently large.

To facilitate numerical computations, we introduce a compact set $\mathcal{X} = \{x \in \mathbb{R}^n: \tilde{H} x \le \tilde{h}\}$ that covers the operating range of the system. This way, the safety requirements in \eqref{equ:ExcluZone} can be represented in the form of $x(t) \in X \cap \mathcal{X}$. We note that in practice, $\mathcal{X}$ can be chosen as a sufficiently large set and only $X$ is used to represent actual safety requirements. 
We also consider a collection of polyhedral sets $\{\tilde{P}_q\}_{q = 1}^{n_q}$ such that $\tilde{P}_q=P_q\cap \mathcal{X}$ for all $q = 1,...,n_q$. 

With $\mathcal{X}$ and $\tilde{P}_q$, $\tilde{\chi}_{\text{safe},k}$ as the approximation of ${\chi}_{\text{safe},k}$ can be numerically computed using Algorithm~\ref{AlgoSafeSet}. The set operations in Algorithm~\ref{AlgoSafeSet}, including set intersection/union, Minkowski sum $\oplus$, Pontryagin difference $\sim$ and projection of polytopes (line~4), can be computed using the Multi-Parametric Toolbox~3 (MPT3) \cite{MPT3}.
\begin{algorithm}
        \caption{Offline computation of $\tilde{\chi}_{\text{safe},k}$}
         \textbf{Input:} {$A_{q,j},B_{q,j},f_{q,j},E_{q,j},U_q,W^\text{a}_q,\tilde{P}_q\ (\forall q \in \{1,...,n_q\}, \forall j\in \{1,...,n^\text{p}_{q}\}),\tilde{\chi}_{\text{safe},k-1}=\bigcup_{i=1}^{r_{k-1}}\tilde{\chi}_{\text{safe},k-1}^i$}\\
        \textbf{Output:}{$\ \tilde{\chi}_{\text{safe},k}$}
        \label{AlgoSafeSet}
        \begin{algorithmic}[1] 
         \State $\chi_{\text{temp}} \gets \emptyset$
            \For {each mode $q\in \{1,...,n_q\}$}
            \For {each $i\in \{1,...,r_{k-1}\}$}
            \State $R \gets \text{Proj}_x\{(x,u): x\in \tilde{\chi}_{\text{safe},k-1}\cap\tilde{P}_q,u\in U_q,\text{s.t.} 
       \begin{bmatrix}
      \bar{A}_q & \bar{B}_q
     \end{bmatrix}
     \begin{bmatrix}
      x \\ u
     \end{bmatrix}\in
     \prod_{j=1}^{n^\text{p}_q} (\tilde{\chi}^i_{\text{safe},k-1}\sim E_{q,j}W_q^\text{a}-f_{q,j}) \}$
            \State $\chi_{\text{temp}} \gets \chi_{\text{temp}}\cup R$
            \EndFor
        \EndFor
        \State $\tilde{\chi}_{\text{safe},k} \gets \chi_{\text{temp}}$
        \end{algorithmic}
\end{algorithm}
After $\tilde{\chi}_{\text{safe},k'}$ has been computed using Algorithm~\ref{AlgoSafeSet}, the constraints in the RAG online optimization problem \eqref{equ:RAG} are approximated as follows,
\begin{align}
\begin{split}\label{equ:OriginalCtr}
    &A_{q(t)}(w^\text{p}_{q(t)})x(t)+B_{q(t)}(w^\text{p}_{q(t)})u
    +E_{q(t)}(w^\text{p}_{q(t)})w^\text{a}_{q(t)}
  + f_{q(t)}(w^\text{p}_{q(t)})\\ & \in \tilde{\chi}_{\text{safe},k'},\ \forall w^\text{p}_{q(t)} \in W^\text{p}_{q(t)}, \forall w^\text{a}_{q(t)} \in W^\text{a}_{q(t)}, \tilde{\chi}_{\text{safe},k'} = \bigcup_{i=1}^{r_k'}\tilde{\chi}_{\text{safe},k'}^i.
\end{split}
\end{align}
In order to deal with the parametric uncertainties, we further tighten \eqref{equ:OriginalCtr} as follows, 
\begin{align}
\begin{split}\label{equ:tightenCtr1}
    & A_{q(t)}(w^\text{p}_{q(t)})x(t)+B_{q(t)}(w^\text{p}_{q(t)})u
    +E_{q(t)}(w^\text{p}_{q(t)})w^\text{a}_{q(t)}\\
    & + f_{q(t)}(w^\text{p}_{q(t)})\in \tilde{\chi}^i_{\text{safe},k'},\ \forall w^\text{p}_{q(t)} \in W^\text{p}_{q(t)}, \forall w^\text{a}_{q(t)} \in W^\text{a}_{q(t)},\\&\text{for some } i\in \{1,...,r_{k'}\}.
\end{split}
\end{align}
Note that \eqref{equ:tightenCtr1} requires there exists control input $u(t)$ such that $f_{q(t)}(x(t),u(t),W_{q(t)})\subset \tilde{\chi}^i_{\text{safe},k'}\subset \tilde{\chi}_{\text{safe},k'}$ for some $i\in \{1,...,r_k'\}$, which is more strict than the constraints \eqref{equ:OriginalCtr}, where $u(t)$ is required to make $f_{q(t)}(x(t),u(t),W_{q(t)})\subset  \tilde{\chi}_{\text{safe},k'}=\bigcup_{i=1}^{r_k'}\tilde{\chi}_{\text{safe},k'}^i$ for any $i\in \{1,...,r_k'\}$. This way the feasible solution of the optimization problem in \eqref{equ:RAG} with the tightened constraint \eqref{equ:tightenCtr1} satisfies the constraint \eqref{equ:ExcluZone}. Using Lemma 1, we can write \eqref{equ:tightenCtr1} in the form as follows,  
\begin{align}
\begin{split}\label{equ:tightenCtr2}
    & A_{q(t),j}x(t)+B_{q(t),j}u+f_{q(t),j}\in \tilde{\chi}^i_{\text{safe},k'} \sim E_{q(t),j}W^\text{a}_{q(t)},\\ &\text{for all } j=\{1,...,n^\text{p}_{q(t)}\},\text{for some } i\in \{1,...,r_{k'}\}.
\end{split}
\end{align}
Since the Pontryagin difference of a polytope $\tilde{\chi}^i_{\text{safe},k'}$ and a polytope $E_{q(t),j}W^\text{a}_{q(t)}$ is still a polytope \cite{kolmanovsky1998theory}, and can be expressed as follows:
\begin{align}\label{equ:Xsafe_finite_union}
\begin{split}
    &\tilde{\chi}^i_{\text{safe},k'} \sim E_{q(t),j}W^\text{a}_{q(t)} = \left\{x \in \mathbb{R}^n: H^{i}_j x \leq h^{i}_j \right\}.
\end{split}
\end{align}
Therefore, the set-inclusion constraints in \eqref{equ:tightenCtr2} can be expressed using the following set of constraints:
\begin{align}\label{equ:RAGMIQP}
\begin{split}
    &H^{i}_j \left(A_{q(t),j}x(t)+B_{q(t),j}u+f_{q(t),j}\right) \leq h^{i}_j+M(1-\delta_i),\\
    &\forall j\in \{1,...,n^\text{p}_{q(t)}\},  \\[4pt]
     &\sum_{i=1}^{r_{k}} \delta_i \geq 1,\, \delta_i \in \{0,1\}.\,
\end{split}
\end{align}
Hence, the RAG online optimization problem \eqref{equ:RAG} can be solved as a Mixed-Integer Quadratic Programming (MIQP) problem using standard MIQP solvers.

\section{Safe reinforcement learning using RAG}\label{sec:4}
\subsection{Safe Online Learning}\label{sec:41}
\begin{figure}[thpb]
      \centering
      \includegraphics[scale=0.45]{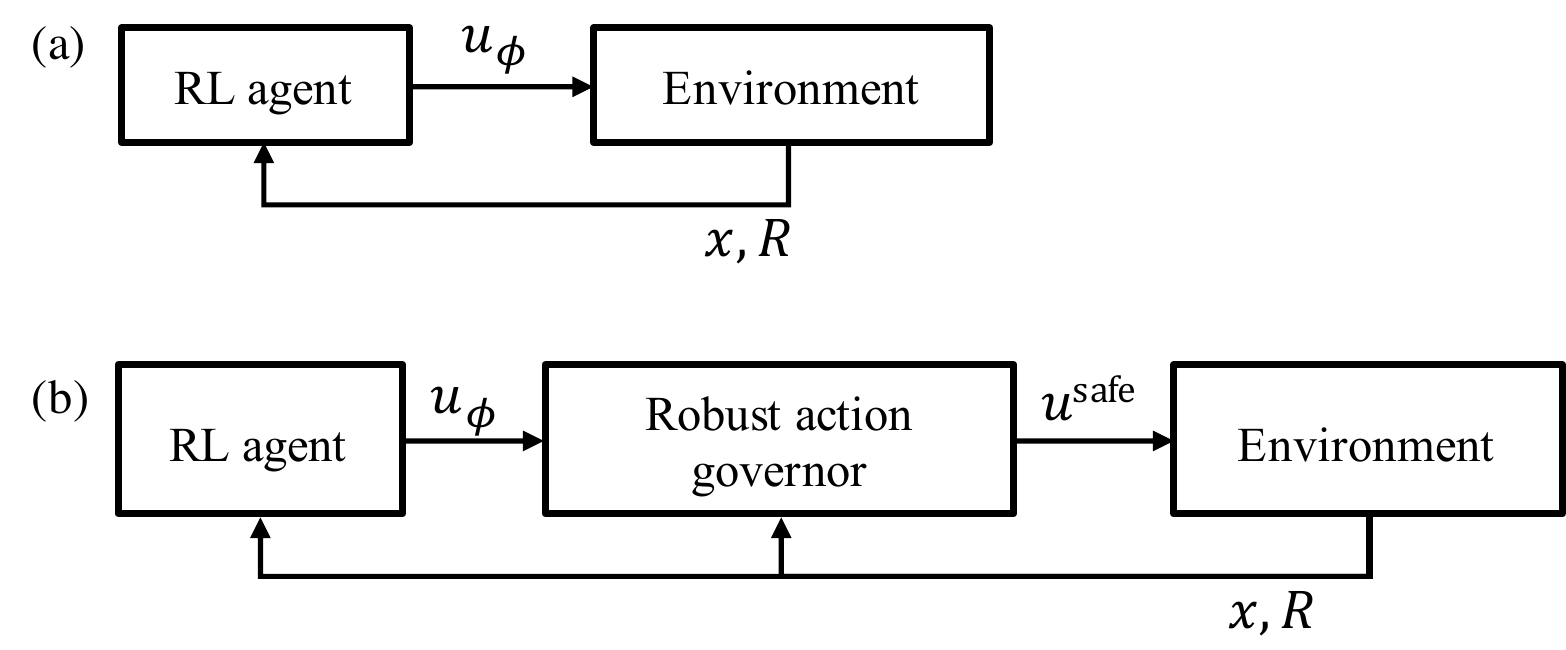}
      \caption{Schematic diagram of conventional RL (a) and safe RL with RAG (b).}
      \label{fig:RL_schematic}
\end{figure}

In this section, we integrate the proposed RAG with a RL algorithm to achieve safe learning that enforces all-time safety constraint satisfaction. The comparison between conventional RL and proposed safe RL scheme is illustrated in Fig. \ref{fig:RL_schematic}. As shown in Fig. \ref{fig:RL_schematic}a, conventional RL agent continuously interacts with the environment to optimize its action via trial-and-error by maximizing its accumulated reward (or minimizing accumulated cost). Specifically, at each time instant $t \in \mathbb{Z}_{\geq 0}$, the agent first takes a measurement of the state $x(t)\in \mathbb{R}^n$, then applies a control action $u_\phi(t)\in \mathbb{R}^m$ based on the current control policy $\pi: \mathbb{R}^n \rightarrow \mathbb{R}^m$, and finally collects a reward $R(t)\in \mathbb{R}$. The control policy $\pi$ is learned and keeps evolving from the collected experience $\{x(t),u_\phi(t),R(t)\}$ to maximize the long-term reward $\tilde{R}=\sum_{t=0}^{\infty} \gamma^t R(t)$, where $\gamma \in (0,1)$ is a discount factor. Apparently, the RL agent needs to explore within the action space to maximize the long-term reward, which may cause unsafe behaviors (i.e., violation of certain safety constraints) during the learning process. This feature hinders the application of RL to online policy learning/evolving of actual engineering systems, which motivates us to propose a safe RL framework.

The proposed safe RL framework is illustrated in Fig. \ref{fig:RL_schematic}b. The RAG is introduced between the RL agent and the environment where the control policy acts on, e.g., the plant in a conventional control setting. This way, the RAG is able to monitor the control signal $u_\phi(t)$ generated by the RL agent and corrects the ones that may cause unsafe behaviors, i.e. violate safety constraints, to the safe action $u^{\text{safe}}(t)$.
As the RAG is an add-on module, no restriction is imposed for the base RL module to be combined with the RAG to achieve safe RL. In this paper, we consider the Neural-Fitted Q-learning (NFQ) as the base RL module to demonstrate the ability to achieve safe learning with RAG.

In NFQ, we use a neural network to approximate the Q-function, and we update this Q-function approximation using the following formulas,
\begin{subequations}
\begin{align}
\label{equ:Q_learning}
 Q(x(t),u(t))\leftarrow & \lambda \tilde{Q}(x(t),u(t)) + (1-\lambda) \big(R(x(t),u(t)) \nonumber \\&+ \gamma \tilde{V}(x(t+1))\big),
\end{align}
\begin{equation}
\label{equ:Q_learning_V}
 \tilde{V}(x) = \max_{u \in {U}} \tilde{Q}(x,u),
\end{equation}
\end{subequations}
where $\lambda \in (0,1)$ is the learning rate, $\gamma \in (0,1)$ is the discount factor, and $\tilde{Q}(x,u)$ represents the approximated Q-values extracted from the neural network. We formalize the proposed safe RL scheme in Algorithm \ref{RL_RAG}. An $\epsilon$-greedy action selection rule is used by the RL agent to balance exploration and exploitation in lines~7-11. The control action is modified in line 12 where any action that may lead to constraint violation will be modified to $u^{\text{safe}}$ by the RAG via solving (\ref{equ:RAG}). The reward is collected in line 14 and the Q-value is updated in lines 15-17, new experience is stored into the replay buffer $B$ in line 18, which is used for training the Q-function NN $\tilde{Q}(x,u)$ in line 23. In particular, in line 17 the Q-value of the current state $x(t)$ and nominal control $u(t)$, $Q(x(t),u(t))$ is updated with the reward $R(x(t),u^{\text{safe}}(t))$ brought by the safe action $u^{\text{safe}}(t)$. Furthermore, in line 18, we store the tuple of current state, nominal control (not the safe action $u^{\text{safe}}(t)$) and updated Q-value, $(x(t),u(t),Q(x(t),u(t)))$, to the replay buffer. This way, the agent is able to explore all possible actions in the action space $U$ as action modification by the RAG is not perceived by the agent. This may be beneficial as this will potentially improve convergence to the optimal control policy \cite{jaakkola1994convergence}. 
\begin{algorithm}
	\caption{Safe RL algorithm}
	\hspace*{\algorithmicindent} \textbf{Input} {Initialized Q-value NN $\tilde{Q}(x,u)$, empty replay buffer $B$, and the maximum trajectory number within a training episode $N$.}
	\begin{algorithmic}[1]
		\For {each training episode}
		\State $n_t \leftarrow 0$
		\While{$n_t < N$}
		\State{Randomly generate an initial state $x(0)\in \chi_{\text{safe}}$}
		\State $t \leftarrow 0$
		    \While{$t<T$}
                \If{$rand()<\epsilon$}
                    \State $u(t)$ takes a random value within $U$
                \Else
                    \State {$u(t)\in\argmax_{u\in U} \tilde{Q}(x(t),u)$}
                \EndIf
                \State{Solve (\ref{equ:RAG}) to modify $u(t)$ to $u^{\text{safe}}(t)$}
                \State{Apply $u^{\text{safe}}(t)$ to system}
                 \State{Observe next state $x(t+1)$ and reward $R(x(t),u^{\text{safe}}(t))$}
                 \State{Update Q value:}
                 \State $\tilde{V}(x(t+1)) = \max_{u\in U}\tilde{Q}(x(t+1),u)$
                 \State{$Q(x(t),u(t))\leftarrow \lambda \tilde{Q}(x(t),u(t)) + (1-\lambda)\big(R(x(t),u^{\text{safe}}(t)) + \gamma \tilde{V}(x(t+1))\big)$}
                \State{Store $(x(t),u(t),Q(x(t),u(t)))$ to replay buffer $B$}
                \State $t \leftarrow t+1$
		    \EndWhile
		    \State $n_t \leftarrow n_t+1$
		    \EndWhile
		    \State{Train Q-value NN $\tilde{Q}(x,u)$ using sampled data from replay buffer $B$}
		\EndFor
	\end{algorithmic} 
	\label{RL_RAG}
\end{algorithm} 
\subsection{Explicit safe RL policy}\label{sec:42}
The safe RL policies obtained from Algorithm \ref{RL_RAG} require implementing together with the RAG to guarantee safety, i.e., numerically solving on-line MIQP optimization problem \eqref{equ:RAG} at each time instant $t$, based on the value of the current state $x(t)$ and the nominal control action selected by the safe RL policy $u_\phi(t)$. The computational cost of solving MIQP problems has been progressively reduced by recent advances in microcontroller and optimization algorithm \cite{bemporad2018numerically,marcucci2020warm,jerez2013embedded}. However, solving online optimization problem still prevents the application of the RAG in many contexts, especially in safety critical applications, where the software certification may be required and conducting one with optimization solver is nontrivial.

The idea of the explicit safe RL policy is to learn the mapping from all states $x(t)\in \chi_\text{safe}$ to $u^\text{safe}(t)$, which is solved by \eqref{equ:RAG}, via imitation learning \cite{pomerleau1991efficient}. In particular, we define a policy as a map from states $x(t)$ to the safe action $u^\text{safe}(t)$, i.e.
\begin{equation}\label{equ:safeRLpolicyDef}
\pi:x(t)\mapsto u^\text{safe}(t).
\end{equation}
This map is determined by numerically solving \eqref{equ:RAG}. In what follows, we pursue an explicit approximation of $\pi$, denoted by $\tilde{\pi}$, exploiting the approach called imitation learning.

The objective of imitation learning is to imitate expert's behavior via learning a control policy from expert's behavior. In this paper, we treat the map $\pi$ as the expert. The imitation learning can be formulated as the following standard supervised learning problem, 
\begin{equation}\label{equ:ImitationLearning}
\tilde{\pi}\in \argmin_{\pi_\theta} \mathbb{E}_{\bar{\bold{s}}\sim\mathbb{P}(\bar{\bold{s}}|\pi)}[\mathcal{L(\pi(\bar{\bold{s}}),\tilde{\pi}(\bar{\bold{s}}})],
\end{equation}
where $\bar{\bold{s}}$ represents the pair $(x,u^\text{safe})$, $\pi_\theta$ presents a policy parameterized by the optimization variable $\theta$ (e.g., the weights of a neural network), $\mathcal{L}$ is a loss function penalizing the difference between expert's policy $\pi$ and its approximation $\tilde{\pi}$. The notation $\mathbb{E}_{\bar{\bold{s}}\sim\mathbb{P}(\bar{\bold{s}}|\pi)}(\cdot)$ is defined as
\begin{equation}\label{equ:Expection}
\mathbb{E}_{\bar{\bold{s}}\sim\mathbb{P}(\bar{\bold{s}}|\pi)}(\cdot)= \int (\cdot) \bold{d}\mathbb{P}(\bar{\bold{s}}|\pi).
\end{equation}
Note that the expectation in procedure \eqref{equ:ImitationLearning} is with respect to the probability distribution $\mathbb{P}(\bar{\bold{s}}|\pi)$ of the data $\bar{\bold{s}}$ that is determined by the expert policy $\pi$, which is essentially the empirical distribution of $\bar{\bold{s}}$ in the pre-collected dataset.
The training of $\tilde{\pi}$ can be conducted either offline or online, for instance, a neural network based $\tilde{\pi}$ may evolve over time by collecting data $(x(t),u^\text{safe}(t))$ online. 

The key feature of applying $\tilde{\pi}$ as a control policy is that the computational cost will be reduced by avoiding solving MIQP in \eqref{equ:RAG}. Using $\tilde{\pi}$ as a control policy has a drawback in that constraint violation may occur. However, in many applications a slight constraint violation is tolerable \cite{li2021chance}, in such cases, a small error between $\pi$ and $\tilde{\pi}$ may be sufficient enough to ensure the system safety.
\section{Numerical examples}\label{sec:5}
In this section, we apply the proposed RAG, safe RL and explicit safe RL schemes to address a soft-landing control problem of a mass-spring-damper system. We first demonstrate the effectiveness of the RAG enforcing constraint satisfaction with a nominal control policy in Section \ref{sec:51}. Then, the safe RL framework proposed in Section \ref{sec:41} is validated in Section \ref{sec:52}. The performances, e.g. control performances and online computational time, of the explicit safe RL policy are reported in Section \ref{sec:53}.
\subsection{Soft-landing control of a mass-spring-damper system using the RAG}\label{sec:51}
\begin{figure}[thpb]
      \centering
      \includegraphics[scale=0.5]{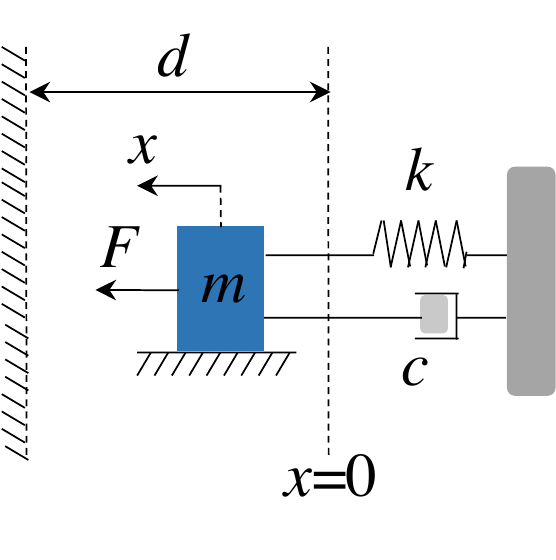}
      \caption{Schematics of a mass-spring-damper system controlled by an external attractive force $F$.}
      \label{fig:MSDSchematic}
\end{figure}

Consider a mass-spring-damper system (MSD) in Fig. \ref{fig:MSDSchematic} where the mass moves within a bounded region with the controlled external force $F$. We impose velocity constraints to the mass when it approaches the boundaries to achieve soft-landing. This soft-landing control of the MSD can be used to model motion control of aircarft systems \cite{rajaram2017laplace}, manipulation of robot arms \cite{nguyen2020active,kraus2020mechatronic} and interaction of mobile robots \cite{urcola2008cooperative}. In particular, in this paper we consider an automated tool-tray transfer systems delivering tools/materials between working stations within an automated assembly line.   

Taking the $x$-axis origin at the neutral position of the spring. The dynamics of the mass-spring-damper system can be modeled as
\begin{subequations}\label{equ:MSD_ContiMdl}
\begin{equation}\label{equ:MSD_ContiMdl1}
m(w^\text{p})\ddot{x}=F-c\dot{x}-F_\text{s}+w^\text{a},
\end{equation}
\begin{equation}\label{equ:MSD_ContiMdl2}
F_\text{s}=\begin{cases}
			k_1x, & \text{if $-x_\text{m}\leq x\leq x_\text{m}$ }\\
            k_2x+(k_1-k_2)x_\text{m}, & \text{$x>x_\text{m}$}\\
            k_2x-(k_1-k_2)x_\text{m}, & \text{$x<-x_\text{m}$}
		 \end{cases}
\end{equation}
\end{subequations}
where $x$ represents the position of the mass $m$ and $c = 0.8\ \text{Ns/m}$ represents the damping coefficient. The nonlinearity of the spring force $F_\text{s}$ is approximated by a piecewise affine model as \eqref{equ:MSD_ContiMdl2}, where $x_\text{m} = 1.75\ \text{m}$ is the spring force mode separation point, $k_1 = 1\ \text{N/m}$ and $k_2 = 0.8\ \text{N/m}$ are spring stiffness for corresponding mode, respectively. In particular, we assume that the controlled external force can only attract but not repel the mass, i.e. $F\geq 0$. This represents the typical scenarios where $F$ is generated by an electromagnetic system \cite{cairano2007model}. $w^\text{a}\in W^\text{a}=\{w^\text{a}\in \mathbb{R}:\ -1\leq w^\text{a}\leq 1\}$ represents a unknown but bounded external control input disturbance. We also assume the mass $m\in [m_\text{min},m_\text{max}]$ is unknown but bounded, taking into account of the fact that the mass of the materials/tools may vary. The uncertainty of $m$ is modeled as parametric disturbances of the system, i.e. $w^\text{p} = [w^\text{p}_1,w^\text{p}_2]^T \in W^\text{p} = \{w^\text{p}_1+w^\text{p}_2=1,w^\text{p}_1,w^\text{p}_2\geq0\}$ and  $\frac{1}{m}=w^\text{p}_1\frac{1}{m^\text{min}}+w^\text{p}_2\frac{1}{m^\text{max}}$.

By using Euler discretization method with sampling time step $T_\text{s}=0.1 \ \text{s}$, equation \eqref{equ:MSD_ContiMdl} becomes
\begin{align}\label{equ:MSD_DiscreteMdl_overall}
\begin{split}
 &z(t+1) =\\
 &\begin{cases}
 \begin{bmatrix}
   1 & T_\text{s} \\
   -w^\text{p}_1\frac{k_1T_\text{s}}{m^\text{min}}-w^\text{p}_2\frac{k_1T_\text{s}}{m^\text{max}} & 1-w^\text{p}_1\frac{cT_\text{s}}{m^\text{min}}-w^\text{p}_2\frac{cT_\text{s}}{m^\text{max}}
   \end{bmatrix}
   z(t) + \\
   \begin{bmatrix} 0 \\  w^\text{p}_1\frac{T_\text{s}}{m^\text{min}}+w^\text{p}_2\frac{T_\text{s}}{m^\text{max}}\end{bmatrix} u(t)
    +\begin{bmatrix}  0  \\ w^\text{p}_1\frac{T_\text{s}}{m^\text{min}}+w^\text{p}_2\frac{T_\text{s}}{m^\text{max}}\end{bmatrix} w^\text{a}(t),\hfill \\\text{if $-x_\text{m}\leq x\leq x_\text{m}$ } \\
    \\
    \begin{bmatrix}
   1 & T_\text{s} \\
   -w^\text{p}_1\frac{k_2T_\text{s}}{m^\text{min}}-w^\text{p}_2\frac{k_2T_\text{s}}{m^\text{max}} & 1-w^\text{p}_1\frac{cT_\text{s}}{m^\text{min}}-w^\text{p}_2\frac{cT_\text{s}}{m^\text{max}}
   \end{bmatrix}
   z(t) + \\
   \begin{bmatrix} 0 \\  w^\text{p}_1\frac{T_\text{s}}{m^\text{min}}+w^\text{p}_2\frac{T_\text{s}}{m^\text{max}}\end{bmatrix} u(t)
    +\begin{bmatrix}  0  \\ w^\text{p}_1\frac{T_\text{s}}{m^\text{min}}+w^\text{p}_2\frac{T_\text{s}}{m^\text{max}}\end{bmatrix} w^\text{a}(t) +\\\begin{bmatrix}  0  \\ -w^\text{p}_1\frac{T_\text{s}(k_1-k_2)x_\text{m}}{m^\text{min}}-w^\text{p}_2\frac{T_\text{s}(k_1-k_2)x_\text{m}}{m^\text{max}}\end{bmatrix}, \hfill \\\text{if $x> x_\text{m}$ }\\
    \\
    \begin{bmatrix}
   1 & T_\text{s} \\
   -w^\text{p}_1\frac{k_2T_\text{s}}{m^\text{min}}-w^\text{p}_2\frac{k_2T_\text{s}}{m^\text{max}} & 1-w^\text{p}_1\frac{cT_\text{s}}{m^\text{min}}-w^\text{p}_2\frac{cT_\text{s}}{m^\text{max}}
   \end{bmatrix}
   z(t) + \\
   \begin{bmatrix} 0 \\  w^\text{p}_1\frac{T_\text{s}}{m^\text{min}}+w^\text{p}_2\frac{T_\text{s}}{m^\text{max}}\end{bmatrix} u(t)
    +\begin{bmatrix}  0  \\ w^\text{p}_1\frac{T_\text{s}}{m^\text{min}}+w^\text{p}_2\frac{T_\text{s}}{m^\text{max}}\end{bmatrix} w^\text{a}(t) +\\\begin{bmatrix}  0  \\ w^\text{p}_1\frac{T_\text{s}(k_1-k_2)x_\text{m}}{m^\text{min}}+w^\text{p}_2\frac{T_\text{s}(k_1-k_2)x_\text{m}}{m^\text{max}}\end{bmatrix}, \hfill \\\text{if $x< -x_\text{m}$ }
 \end{cases}    
\end{split}
\end{align}
where $z=[x,\dot{x}]^T$ and $u=F$.
To avoid high velocity contacts which lead noise and wear, as the mass approaches the boundary ($x=d$), which represents the position of a working station, it is desirable to reduce the velocity of the mass gradually to achieve soft-landing, which can be represented by imposing the following constraint
\begin{align}\label{equ:SoftLandingConstr}
\begin{split}
\begin{cases}
 -\epsilon - \beta (d-x)\leq \dot{x} \leq \epsilon + \beta (d-x),& \text{if $x_\text{c}\leq x\leq d$ }\\
 -v_\text{max}\leq \dot{x} \leq v_\text{max},& \text{if $ x\leq x_\text{c}$ }
\end{cases}
\end{split}
\end{align}
The constraint \eqref{equ:SoftLandingConstr} enforces that when the mass contacts the boundary ($x=d$), the velocity is within $[-\epsilon,\epsilon]$. This range is gradually relaxed by $\beta$ as the mass moves away from the boundary. Note that the soft-landing constraint is not required when $x\leq x_\text{c}$, which leads \eqref{equ:SoftLandingConstr} to be a non-convex constraint in general.

 We also enforce the following constraint to prevent the mass moving too far away from the working station which can be represented as
\begin{align}\label{equ:PositionConstr}
\begin{split}
x \geq 0.
\end{split}
\end{align}

In addition, the constraint on the control input $F$ is defined as 
\begin{align}\label{equ:CtrConstr}
\begin{split}
 0\leq F \leq F_{\text{max}}.
\end{split}
\end{align}

A RL control policy is employed as the nominal control $u_\phi=\phi(x,\dot{x},x^\text{ref})$, the reward function of the RL agent $R$ is designed as follows, 
\begin{subequations}\label{equ:RewardFunc}
\begin{equation}
R=w_1R_1+w_2R_2,
\end{equation}
\begin{equation}
R_1 = -(x-x^\text{ref})^2
\end{equation}
\begin{equation}
R_2 = \begin{cases}
			-(\dot{x}-v_\text{b})^2, & \text{if \eqref{equ:SoftLandingConstr} is violated}\\
            0, & \text{if \eqref{equ:SoftLandingConstr} is satisfied}
		 \end{cases}
\end{equation}
\end{subequations}
where $w_1$ and $w_2$ are tuning weightings. $x^\text{ref}$ is the reference position for the mass, in particular, we assume $x^\text{ref}\in \{0,d\}$. $v_\text{b}$ represents the closest velocity constraint value determined by \eqref{equ:SoftLandingConstr} corresponding to the current position $x$, which can be defined as follows,
\begin{align}\label{equ:ConstrBoundary}
\begin{split}
 v_\text{b} = \begin{cases} -\epsilon - \beta (d-x),& \text{if $x_\text{c}\leq x\leq d$ and  $\dot{x}\leq 0$}\\
 \epsilon + \beta (d-x),& \text{if $x_\text{c}\leq x\leq d$ and  $\dot{x}> 0$}\\
 -v_\text{max},& \text{if $ x\leq x_\text{c}$ and  $\dot{x}\leq 0$}\\
 v_\text{max},& \text{if $ x\leq x_\text{c}$ and  $\dot{x}> 0$}
\end{cases}
\end{split}
\end{align}


The rest of parameters used in the system model and constraints are selected as $d=5\ \text{m}$, $v_\text{max}=5\ \text{m/s}$, $F_\text{max}=10\ \text{N}$, $x_\text{c}=3.3\ [\text{m}]$, $\epsilon=0.5 \ \text{m/s}$, $\beta= -2.95 \ \text{s}^{-1}$, $m_\text{min} = 0.5\ \text{kg}$ and $m_\text{max} = 1.5\ \text{kg}$.

\begin{figure}[thpb]
      \centering
      \includegraphics[scale=0.45]{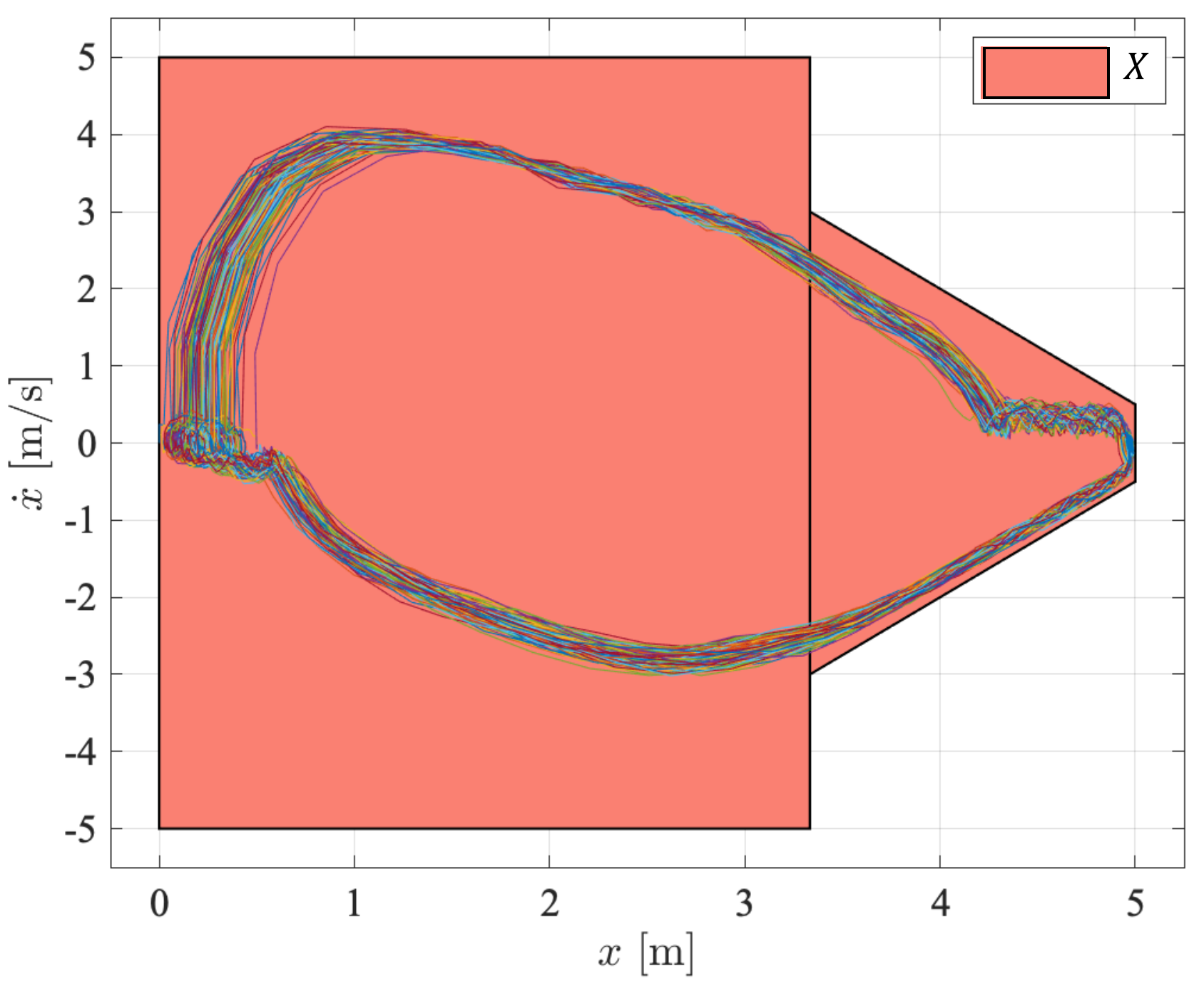}
      \caption{Constraint violation tests over 500 experiments with randomly generated parametric and additive disturbances.}
      \label{fig:StateTrajBeforeRandom}
\end{figure}

Apparently, the satisfactions of constraints \eqref{equ:SoftLandingConstr} and \eqref{equ:PositionConstr} are not guaranteed with the nominal RL control policy $u_{\phi}$. To enforce such safety constraints, the RAG is designed considering the safe set $X$ based on \eqref{equ:SoftLandingConstr} and \eqref{equ:PositionConstr}, which is in general non-convex and shown in Fig. \ref{fig:StateTrajBeforeRandom}. In this paper, $\chi_\text{safe}$ is computed with MPT3 toolbox offline \cite{MPT3}. The online MIQP optimization problem in (\ref{equ:RAG}) is solved by Gurobi \cite{gurobi}.

\begin{figure}[thpb]
      \centering
      \includegraphics[scale=0.45]{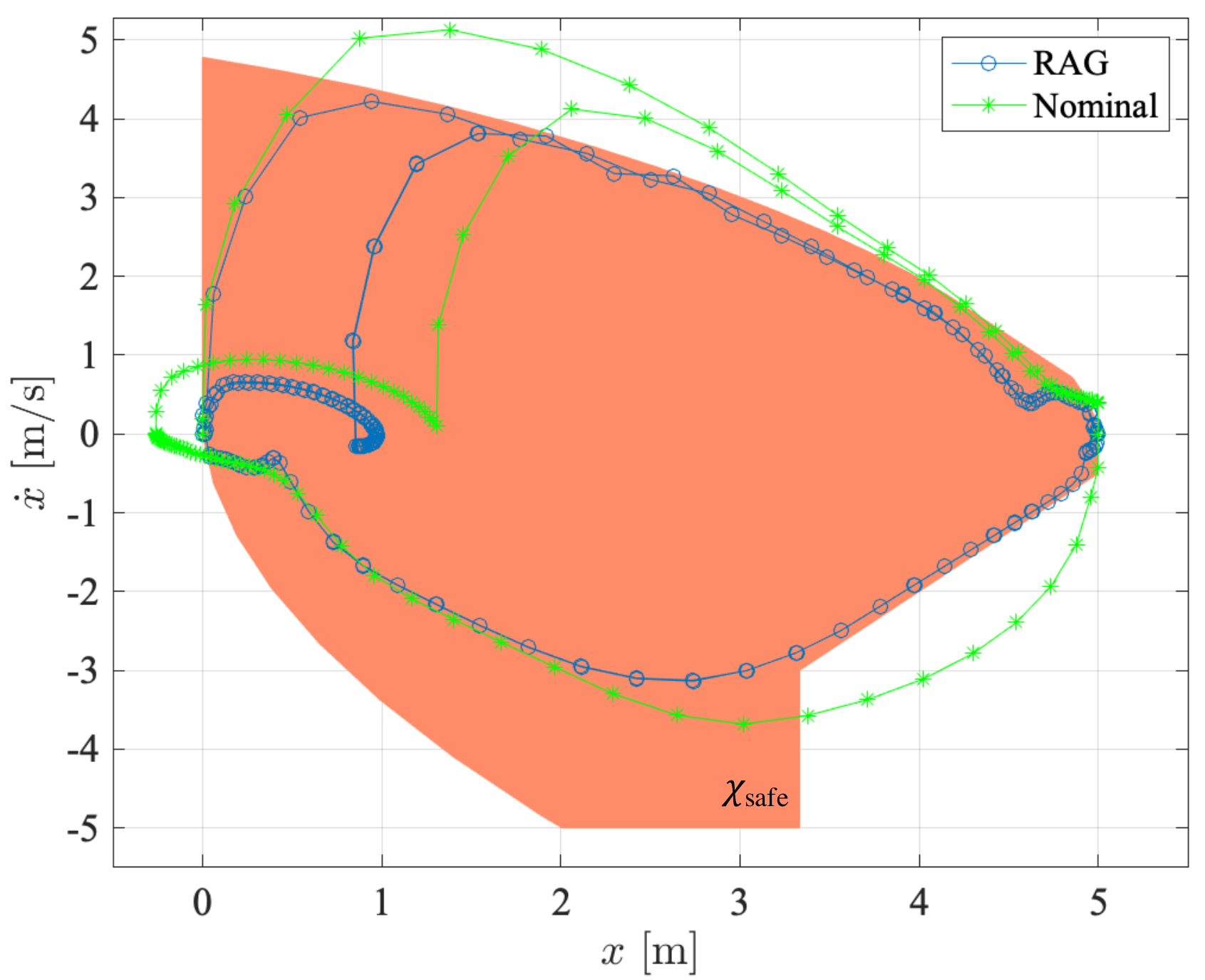}
      \caption{Comparison of state trajectory with the RAG and nominal control policy under adversarial additive disturbances $w^\text{a}$ and computational results of $\chi_\text{safe}$.}
      \label{fig:Xsafe_traj}
\end{figure}

\begin{figure}[thpb]
      \centering
      \includegraphics[scale=0.47]{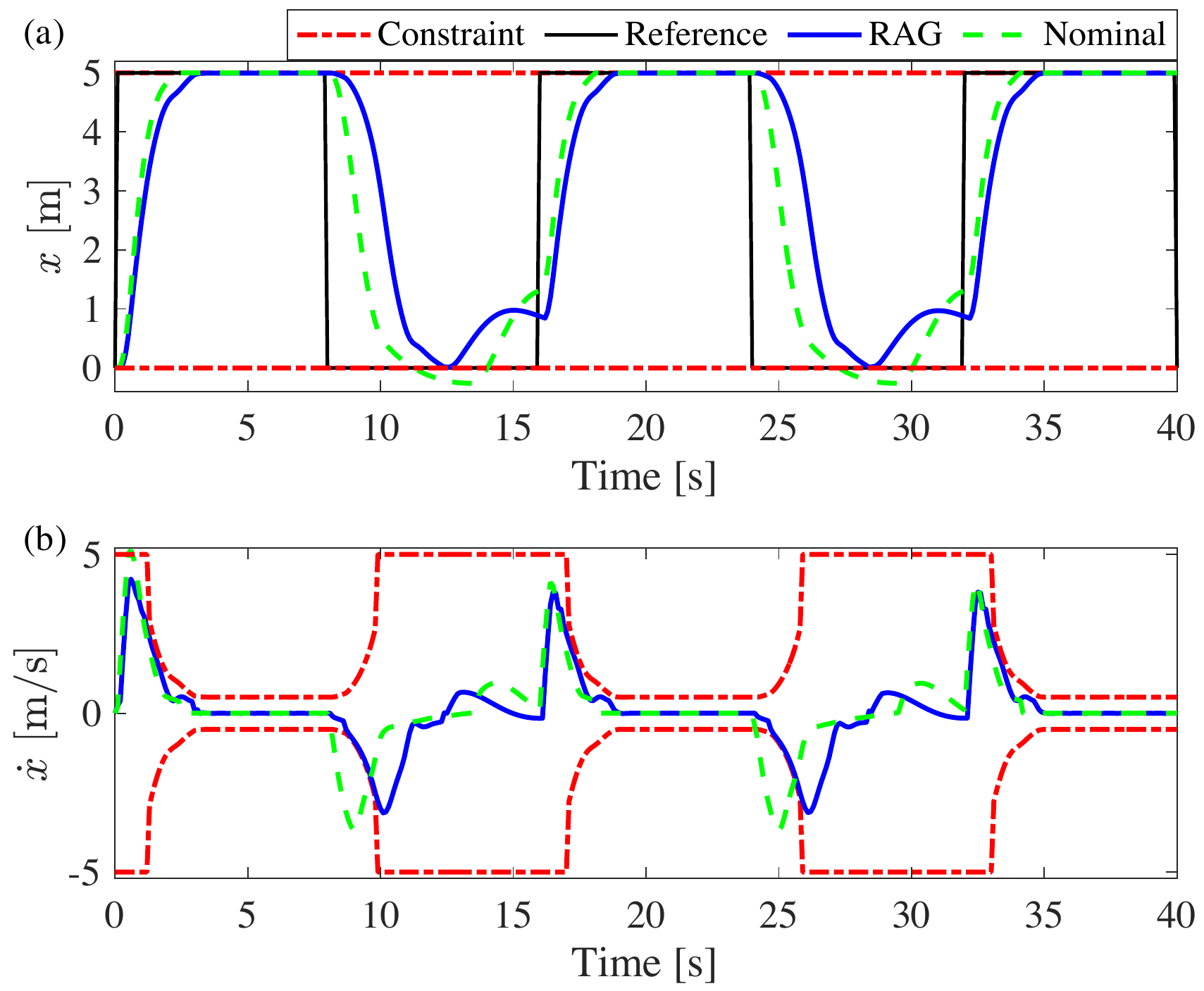}
      \caption{Simulation results of states and constraints with the RAG and nominal control. (a) The displacement of the mass. (b) The velocity of the mass.}
      \label{fig:StateCompareBefore}
\end{figure}

We consider the initial condition $[x(0),\dot{x}(0)]^T=[0,0]^T$ and the desired position of the mass is shown as a periodical signal between 0 and 5 (black solid line) in Fig. \ref{fig:StateCompareBefore}(a). We set $k=60$ and compute the safe set $\chi_\text{safe}$ using Algorithm \ref{AlgoSafeSet}. The set $\chi_\text{safe}$ and state trajectory using the RAG are shown in Fig. \ref{fig:Xsafe_traj}. We can observe that the RAG keeps the state trajectory always within $\chi_\text{safe}$ while the trajectory with the nominal RL control policy gets out of $\chi_\text{safe}$, which leads constraint violations shown in Fig. \ref{fig:StateCompareBefore}.

\begin{figure}[thpb]
      \centering
      \includegraphics[scale=0.48]{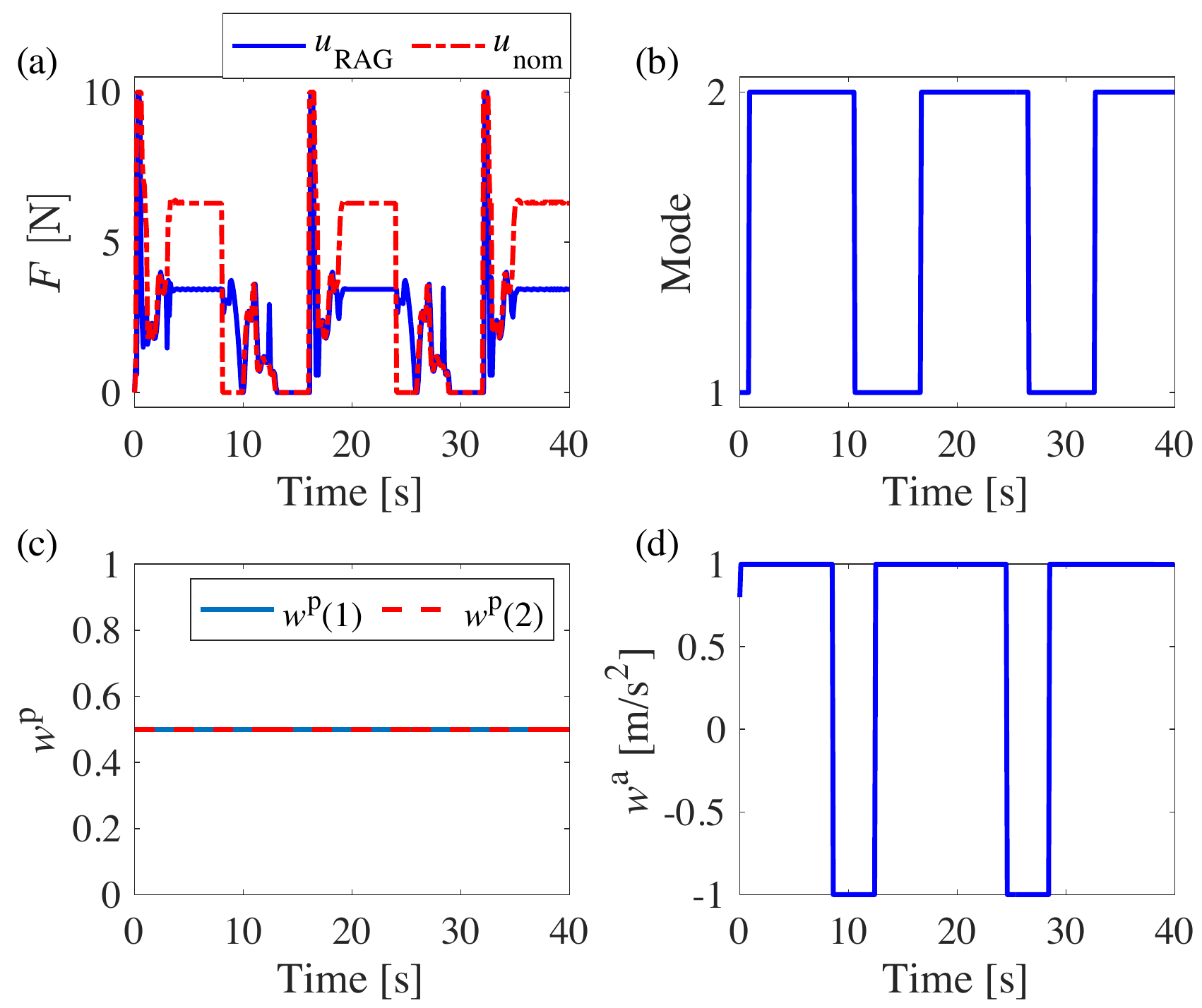}
      \caption{Control inputs comparisons of the RAG and nominal control policy, results of mode transitions and disturbances. (a) Control inputs. (b) Mode. (c) Parametric disturbances. (d) Additive disturbances.}
      \label{fig:CtrDisturBefore}
\end{figure}

The comparisons of states and control inputs using the RAG and that using the nominal RL control policy are presented in Figs. \ref{fig:StateCompareBefore} and \ref{fig:CtrDisturBefore}. The safety constraints in \eqref{equ:SoftLandingConstr} and \eqref{equ:PositionConstr} (red dash-dotted line in Fig. \ref{fig:StateCompareBefore}) are violated with the nominal control policy. In contrast to the results with nominal control, as shown in Figs. \ref{fig:StateCompareBefore} and \ref{fig:CtrDisturBefore}(a), the RAG enforces constraint satisfaction by modifying the nominal control $u_{\phi}$ in advance to decrease the mass velocity when it approaches the boundary ($x=5\  \text{[m]}$) and origin. This way the mass can contact the boundary with a smoother way and also will not move away cross the origin. The mode transition can be safely handled by using the RAG, which is shown in Fig. \ref{fig:CtrDisturBefore}(b). The parametric and additive disturbances applied during simulation are shown in Figs. \ref{fig:CtrDisturBefore}(c) and \ref{fig:CtrDisturBefore}(d). To validate the performances of the RAG, in this test, the adversarial additive disturbances are imposed where the disturbance tries its best to violate constraints, i.e., apply maximum disturbance in the same moving direction of the mass. 

To verify that the RAG can guarantee constraint satisfaction in the presence of any disturbance realizations, we run $500$ simulation experiments with randomized parametric and additive disturbance inputs $w(k) \in W$. The simulated trajectories are shown in Fig.~\ref{fig:StateTrajBeforeRandom}. It can be observed that no constraint violation occurs in these 500 experiments.

\subsection{Safe reinforcement learning with the RAG}\label{sec:52}

In this section, we demonstrate the effectiveness of the proposed safe RL framework by applying it to achieve safe online training for the nominal control policy used in the last section. We assume that the parameter value of the actual mass $m$ and distance between the origin and boundary $d$ are different from the ones used to design the nominal RL control policy $\phi$. This is common in the real implementations as either $m$ and $d$ are difficult to measure precisely or their values vary during the system operation. Our goal is to online evolve the control policy $\phi$ safely, i.e., without violating constraints \eqref{equ:SoftLandingConstr} and \eqref{equ:PositionConstr}, to adapt to the new parameters $m=1.3\ \text{kg}$ and $d=5.5\ \text{m}$ via RL.

\begin{figure}[thpb]
      \centering
      \includegraphics[scale=0.47]{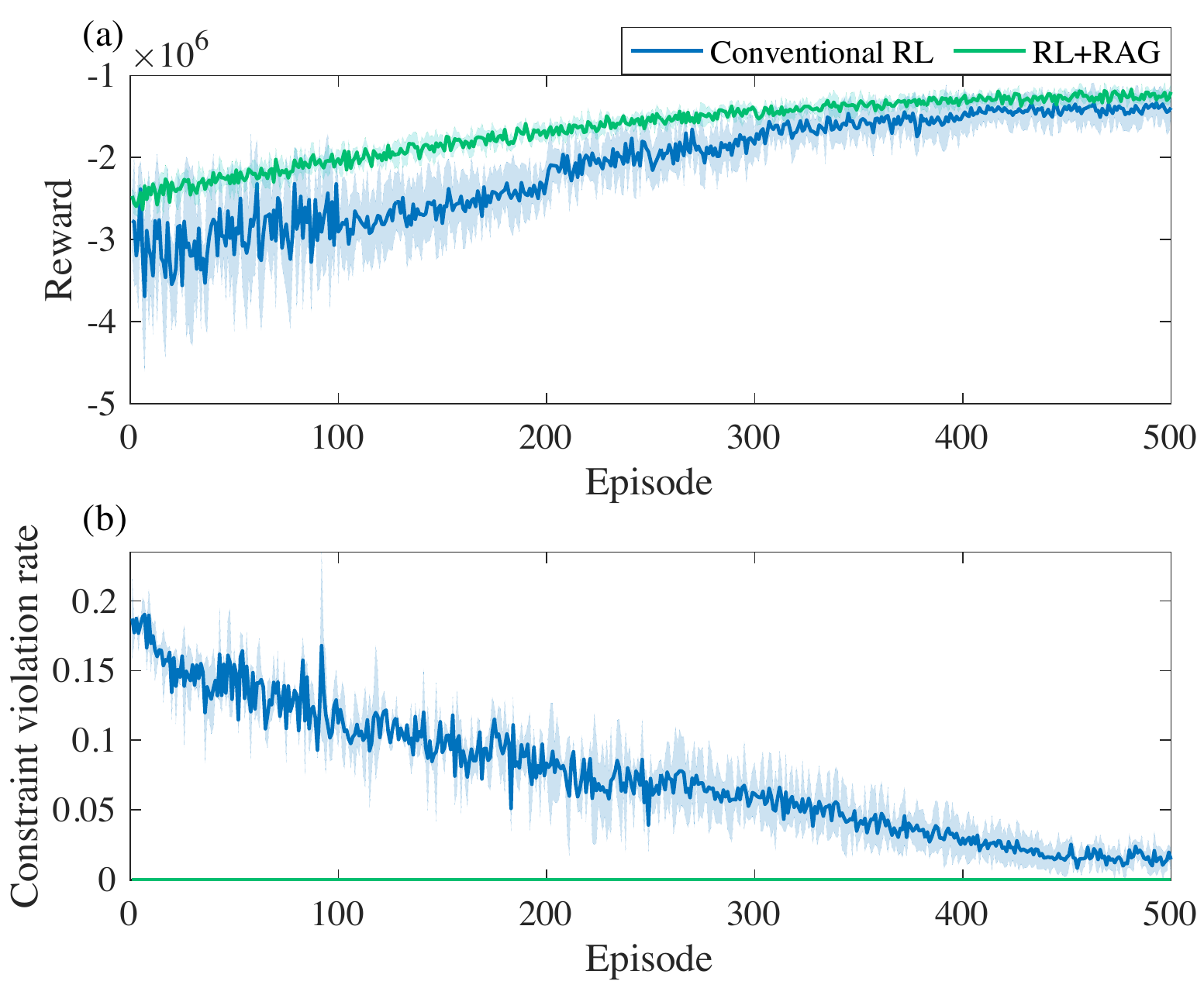}
      \caption{Training histories of conventional RL and safe RL. (a) Constraint violation rates of each episode. (b) Average reward values of each episode. Solid lines represent average values and shaded areas represent the standard deviation values over 20 experiments.}
      \label{fig:RLRewardCV}
\end{figure}

We use the same reward function in \eqref{equ:RewardFunc} to conduct the online training. The comparison between training histories of conventional and safe RL algorithms are shown in Fig. \ref{fig:RLRewardCV}. We can observe from \ref{fig:RLRewardCV}(b) that the constraint violation rate of conventional RL keeps decreasing as the learning proceeds. In contrast, no constraint violation is exhibited for the safe RL with the RAG during the entire training process. Moreover, by using the RAG, the second term in the reward function $R_2$ is always equal to zero, which leads to smaller expectation and variation  values compared with the ones with the conventional RL, as shown in Figure \ref{fig:RLRewardCV}(a). 

\begin{figure}[thpb]
      \centering
      \includegraphics[scale=0.45]{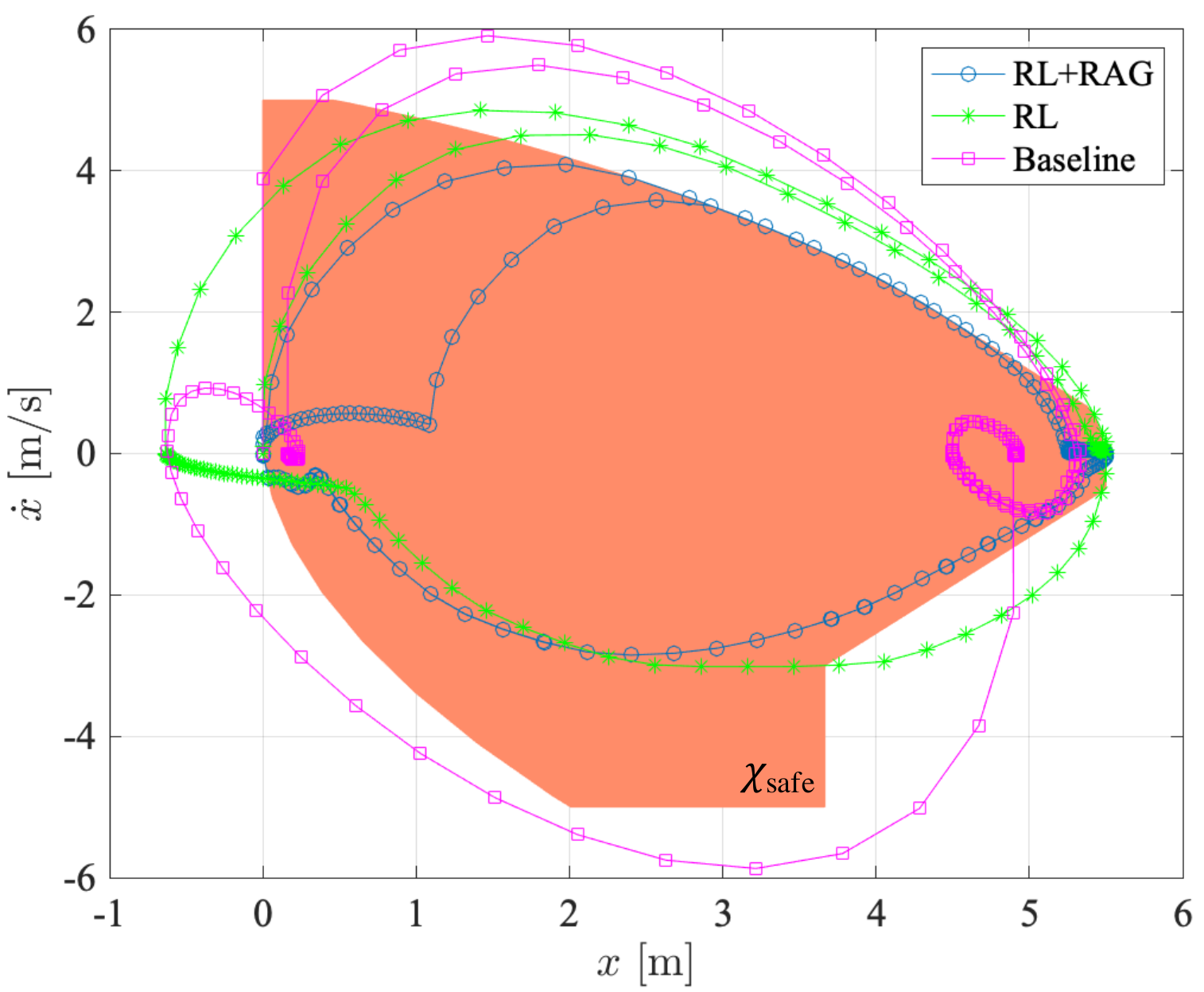}
      \caption{Comparisons of trajectories with controls of baseline RL control policy (Baseline), updated RL control policy (RL) and updated RL control policy with the RAG (RL+RAG).}
      \label{fig:StateTrajAfter}
\end{figure}

\begin{figure}[thpb]
      \centering
      \includegraphics[scale=0.47]{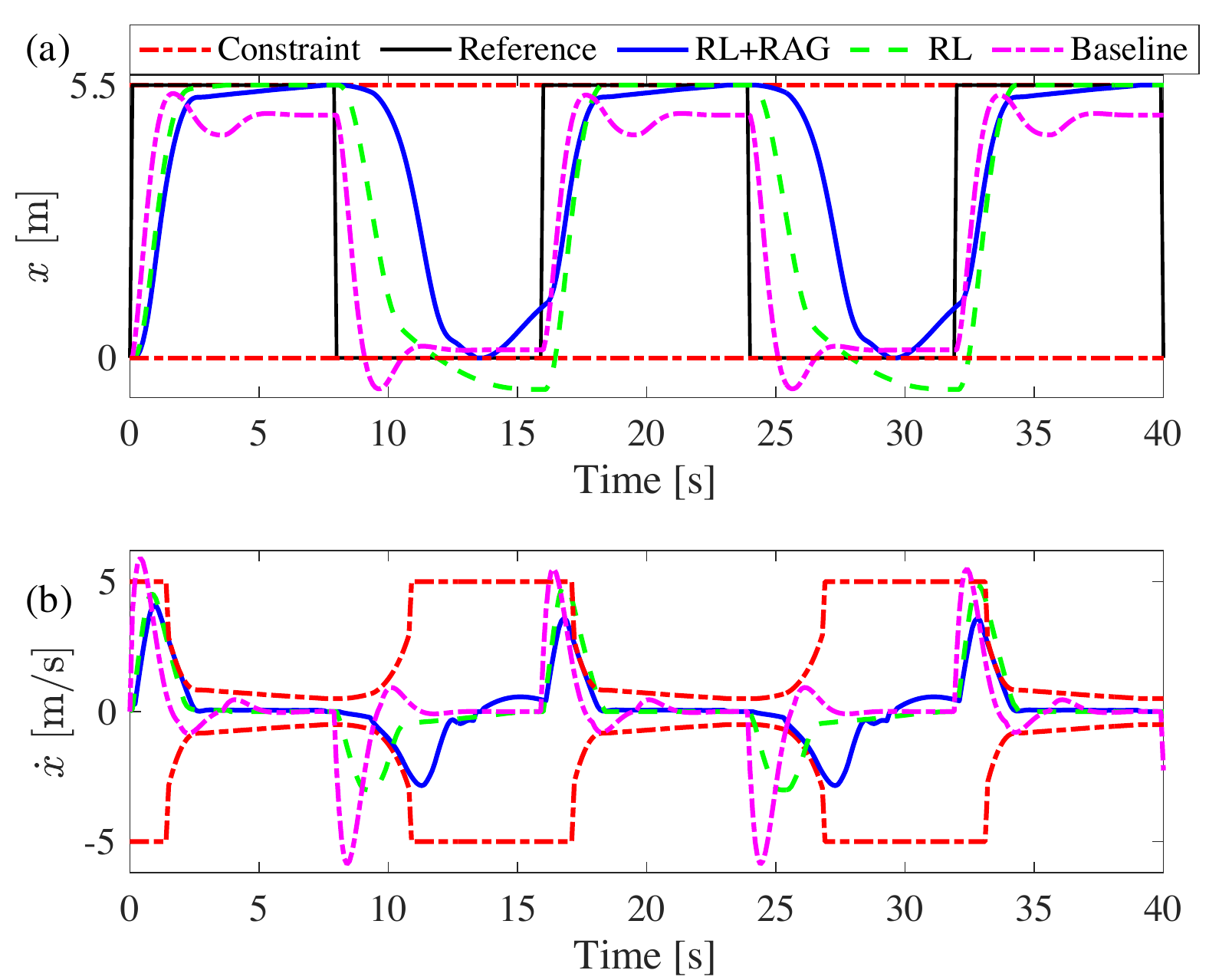}
      \caption{Simulation results of states and constraints with the RAG. (a) The displacement of the mass. (b) The velocity of the mass.}
      \label{fig:StateCompareAfter}
\end{figure}

\begin{figure}[thpb]
      \centering
      \includegraphics[scale=0.48]{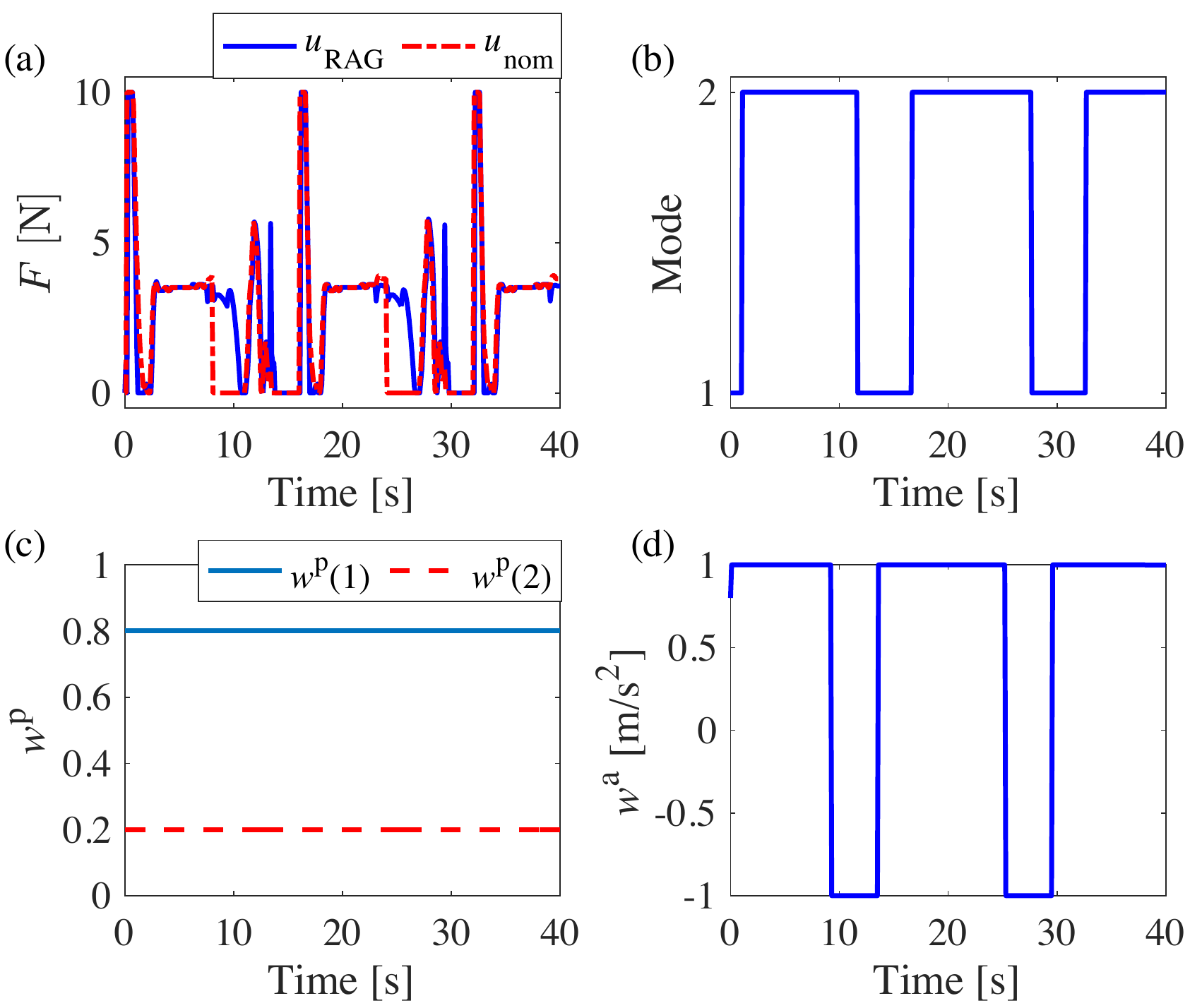}
      \caption{Control inputs comparisons of the RAG and nominal control policy, results of mode transitions and disturbances. (a) Control inputs. (b) Mode. (c) Parametric disturbances. (d) Additive disturbances.}
      \label{fig:CtrDisturAfter}
\end{figure}

The validation results of trained policies are shown in Figs. \ref{fig:StateTrajAfter} to \ref{fig:CtrDisturAfter}. Although the tracking performances of reference are both satisfactory with the conventional RL (green dash line in Fig. \ref{fig:StateCompareAfter}) and safe RL (blue solid line in Fig. \ref{fig:StateCompareAfter}), constraint satisfaction is enforced by the control policy with the RAG, while under the control of conventional RL policy, there are still occasional constraint violations. This can also be verified by Fig. \ref{fig:StateTrajAfter} as the state trajectory is always within $\chi_\text{safe}$. Similar results as in section \ref{sec:51} of control inputs and mode transitions can be observed in Fig. \ref{fig:CtrDisturAfter}, where the RAG modifies the nominal control actions preventing constraint violation from happening.  

We remark that the fact that our safe RL algorithm based on the use of the RAG guarantees no safety constraint violation during both the training and the operating phases yields that it can be used for onboard applications. For instance, it can be used to continuously improve the performance of a controller during its onboard operation.

\subsection{Results of explicit safe RL policy}\label{sec:53}
\begin{figure}[thpb]
      \centering
      \includegraphics[scale=0.67]{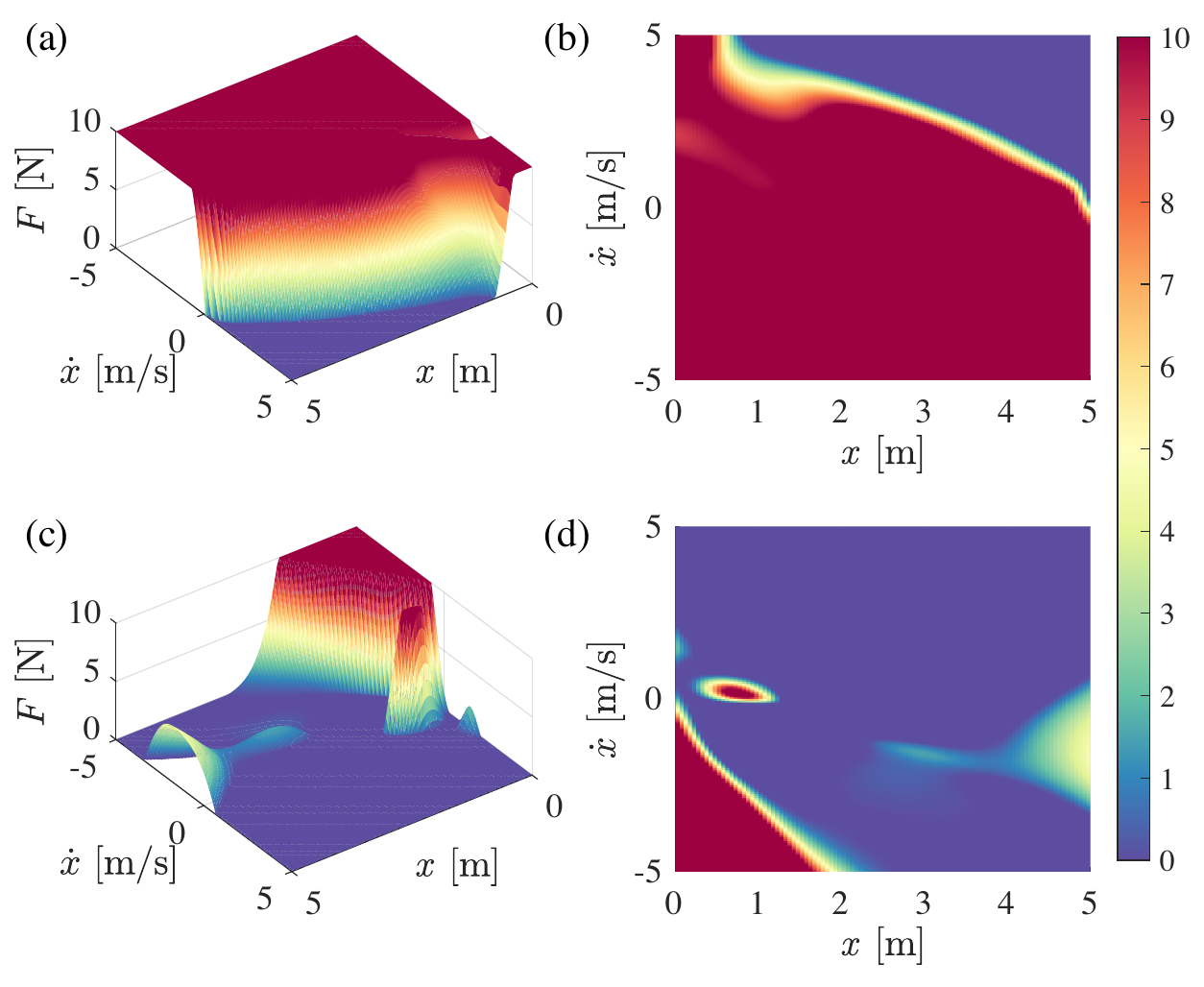}
      \caption{ERAG control policy. (a)-(b) Control policy with $x^\text{ref}=5$. (c)-(d) Control policy with $x^\text{ref}=0$. }
      \label{fig:ERAG_policy}
\end{figure}

In this section, we compare the performances of the explicit safe RL policy \eqref{equ:ImitationLearning} and the conventional RAG \eqref{equ:RAG} using the example in Section \ref{sec:51}. A trained RL policy together with the RAG is served as expert. We use a neural network to approximate the expert's action, which is the exact mapping between states $x$ and safe control inputs $u^\text{safe}$ obtained by solving the RAG optimization problem \eqref{equ:RAG}, via the imitation learning in \eqref{equ:ImitationLearning}. After training, we apply the explicit safe RL policy as shown in Fig. \ref{fig:ERAG_policy} directly to the MSD system to track reference mass position while trying to keep the constraints \eqref{equ:SoftLandingConstr} being satisfied. 

\begin{figure}[thpb]
      \centering
      \includegraphics[scale=0.67]{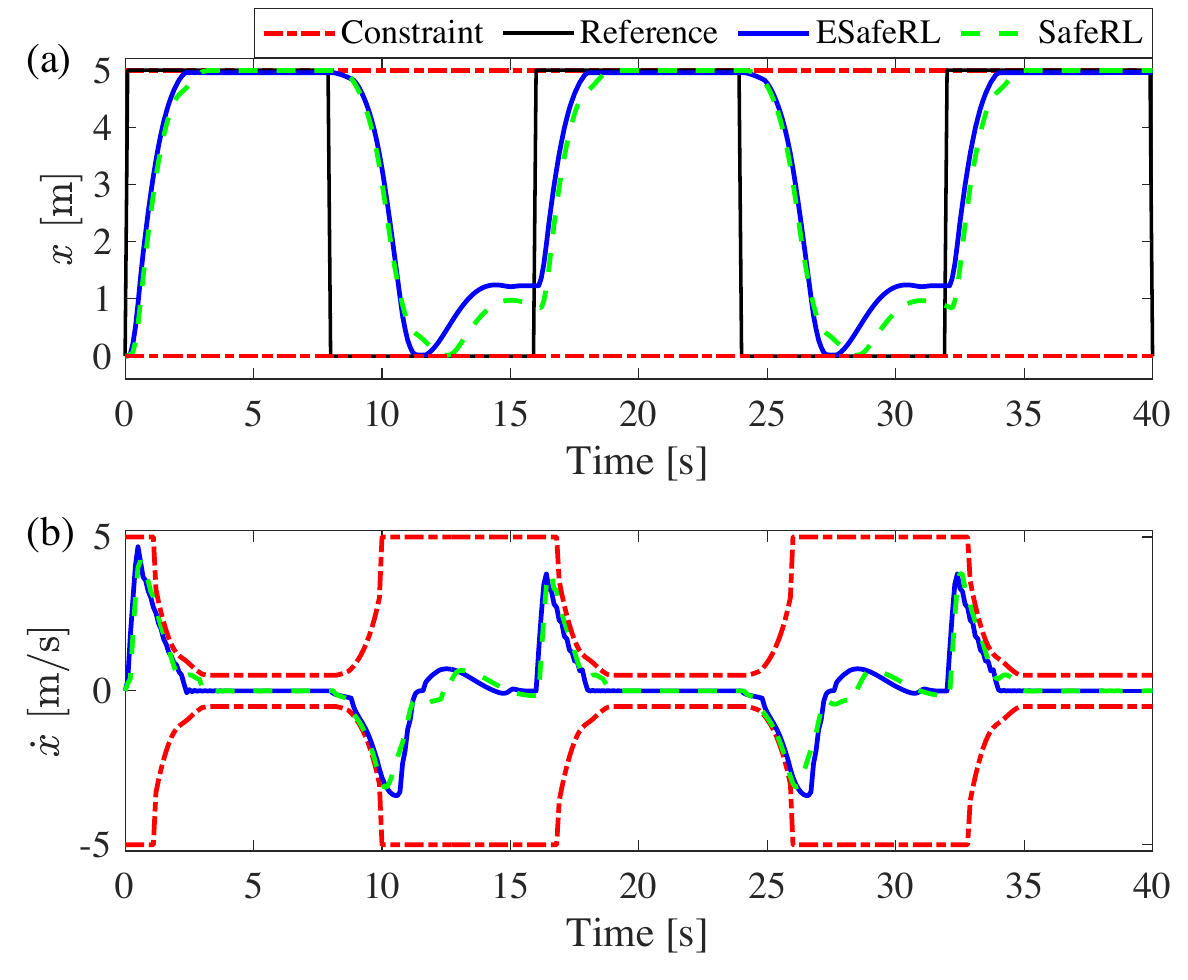}
      \caption{Simulation results of states and constraints with the ERAG and RAG. (a) The displacement of the mass. (b) The velocity of the mass.}
      \label{fig:StateCompareERAG}
\end{figure}

\begin{figure}[thpb]
      \centering
      \includegraphics[scale=0.68]{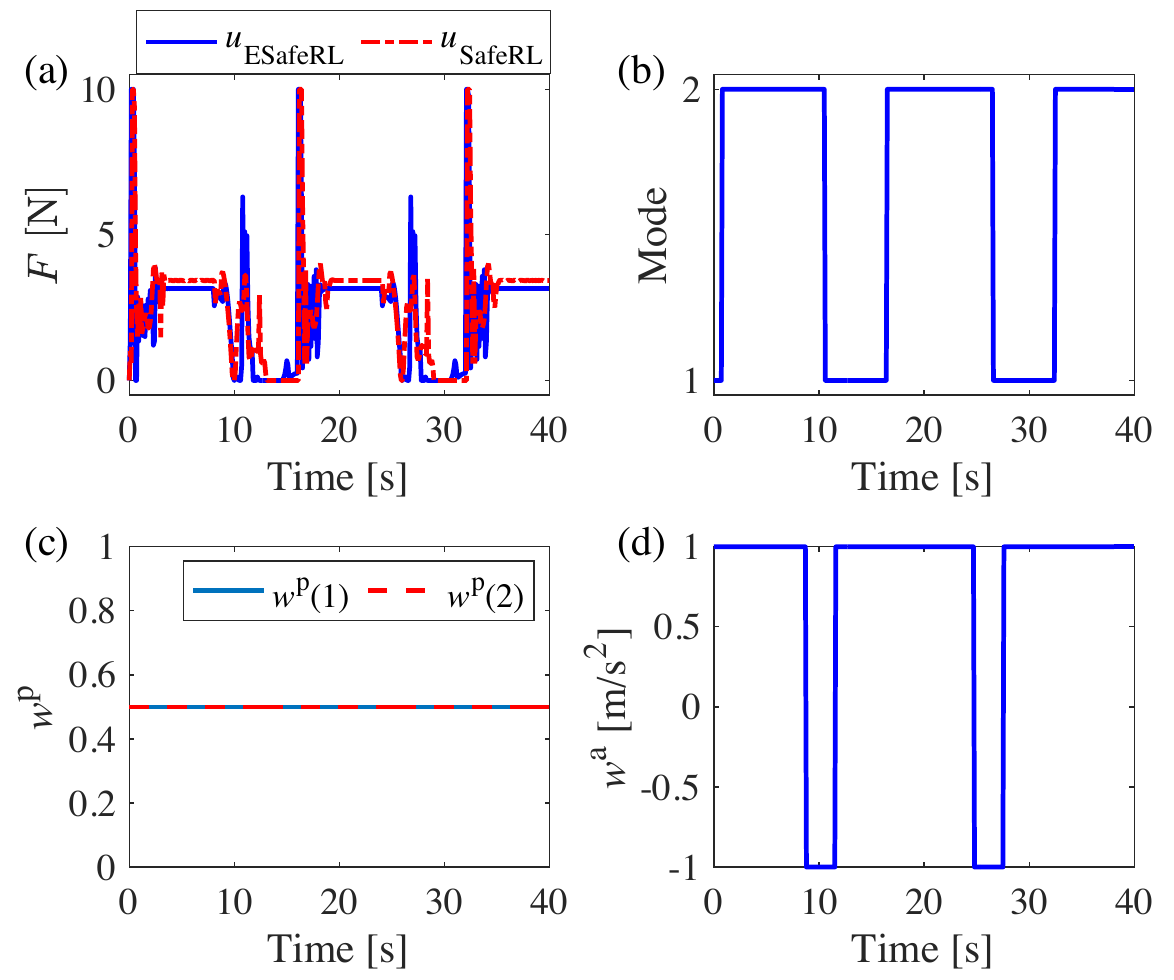}
      \caption{Control inputs comparisons of the ERAG and RAG, results of mode transitions and disturbances. (a) Control inputs. (b) Mode. (c) Parametric disturbances. (d) Additive disturbances.}
      \label{fig:CtrDisturERAG}
\end{figure}

\begin{figure}[thpb]
      \centering
      \includegraphics[scale=0.67]{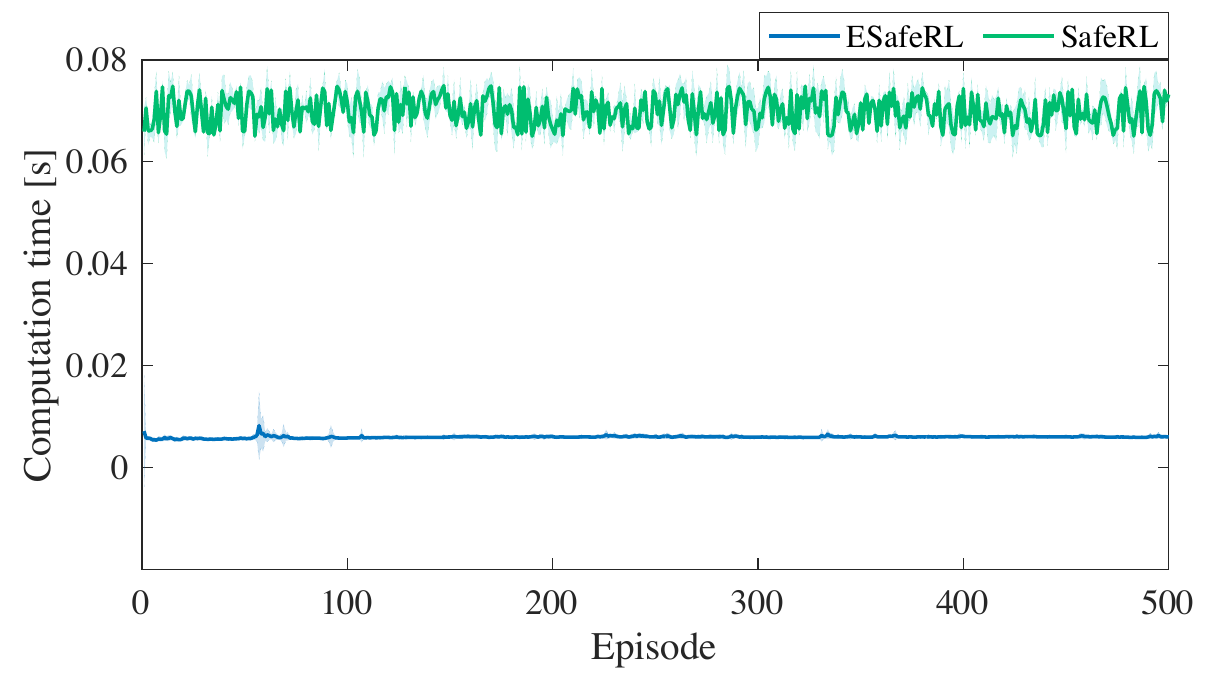}
      \caption{Comparisons of online computational time between explicit safe RL and safe RL policy over 500 experiments with randomly generated initial conditions and disturbances. The solid line represents the average value and shaded area represents the standard deviation value.}
      \label{fig:ComputeTime}
\end{figure}

The control performances using the explicit safe RL policy (ESafeRL) and safe RL policy with the RAG (SafeRL) are compared in Fig. \ref{fig:StateCompareERAG} and \ref{fig:CtrDisturERAG}. Similar to the SafeRL, the mass tracks the reference position and achieves soft-landing  with the control of ESafeRL. Meanwhile, as reported in Fig. \ref{fig:ComputeTime}, the ESafeRL reduces the online computational time by over 95\% on average compared with one with SafeRL by avoiding solving MIQP \eqref{equ:RAG} online.   

\section{Conclusions}\label{sec:6}
In this paper, we introduced an extension of an add-on scheme referred to as the Robust Action Governor (RAG), to the case where discrete-time piecewise affine (PWA) models subject to both additive and parametric uncertainties and non-convex constraints are treated. The RAG modifies the nominal control input when it becomes necessary to enforce constraints. Theoretical properties of robust constraint satisfaction and recursive feasibility have been established. A safe reinforcement learning (RL) framework is established based on the RAG. Applications of the RAG and safe RL to a soft-landing control problem of a mass-spring-damper system have been successfully demonstrated in simulations. Future work will focus on extensions of the RAG to systems with state-dependent uncertainties and on computation efficiency improvements.

\begin{ack}                               
This work is supported by Ford Motor Company. 
\end{ack}

\bibliographystyle{plain} 
\bibliography{reference.bib}{}

\end{document}